\documentclass[prb, twocolumn, showpacs, amsmath, amssymb,superscriptaddress]{revtex4-1}
\usepackage{mysetup}
\usepackage{braket}
\usepackage{bbm}

\newcommand{\la}{\Lambda}
\newcommand{\tn}{\textnormal}

\begin{document}
\title{Second-order functional renormalization group approach\\ to quantum wires out of equilibrium}

\author{C. Kl\"ockner}
\affiliation{Technische Universit\"at Braunschweig, Institut f\"ur Mathematische Physik, Mendelssohnstraße 3, 38106 Braunschweig, Germany}

\author{D.M.\ Kennes}
\affiliation{Institut f\"ur Theorie der Statistischen Physik, RWTH Aachen University and JARA-Fundamentals of Future Information Technology, 52056 Aachen, Germany}
\affiliation{Max  Planck  Institute  for  the  Structure  and  Dynamics  of  Matter,Luruper  Chaussee  149,  22761  Hamburg,  Germany}

\author{C.\ Karrasch}
\affiliation{Technische Universit\"at Braunschweig, Institut f\"ur Mathematische Physik, Mendelssohnstraße 3, 38106 Braunschweig, Germany}

\begin{abstract} 
The functional renormalization group (FRG) provides a flexible tool to study correlations in low-dimensional electronic systems. In this paper, we present a novel FRG approach to the steady-state of quantum wires out of thermal equilibrium. Our method is correct up to second order in the two-particle interaction and accounts for inelastic scattering. We combine semi-analytic solutions of the flow equations with MPI parallelization techniques, which allows us to treat systems of up to 60 lattice sites. The equilibrium limit is well-understood and serves as a benchmark. We compute effective distribution functions, the local density of states, and the steady-state current and demonstrate that all of these quantities depend strongly on the choice of the cutoff employed within the FRG. Non-equilibrium is plagued by the lack of physical arguments in favor of a certain cutoff as well as by the appearance of secular higher-order terms which are only partly included in our approach. This demonstrates the inadequacy of a straightforward second-order FRG scheme to study interacting quantum wires out of equilibrium in the absence of a natural cutoff choice.

\end{abstract}

\date{\today} 
\maketitle

\section{Introduction}

A system that is driven out of equilibrium by a quantum quench will display many interesting, non-equilibrium phenomena (for a review, see Ref.~\onlinecite{Mitra2018}). However, it is generally believed that after some transient period, a generic, infinite (or in practice large) system equilibrates; time-translational invariance is recovered and all local observables are well described by a thermal distribution.\cite{DAlessio2016,Srednicki1999,Neumann1929,Goldstein2010}

A simple way to prevent thermalization is continuous driving of the system. In such a scenario, however, energy is not conserved, and the unique stationary state of a generic, ergodic system is thus an infinite-temperature state.\cite{Hone1997,Lazarides2014, DAlessio2014} Well-known exceptions are integrable models (e.g., non-interacting ones)\cite{Lazarides2014b} as well as many-body localized systems,\cite{Ponte2015} which display emergent integrability.\cite{Ros2015,Imbrie2016, Imbrie2016b,Imbrie2017}

One way to circumvent heating is to consider an infinite (i.e., open) quantum system where an infinite amount of energy can dissipate. Such a setup is attractive from a theoretical perspective: it prevents recurrence, allows for a non-trivial stationary state even without time-dependent driving, and is physically-relevant as some coupling to the environment can never be fully eliminated. Treating an open quantum system, however, presents a major hurdle to many theoretical methods. 
It is a priori unclear how long transient dynamics persist and when a stationary state will be reached. Additionally, effects like prethermalization\cite{Berges2004} render the use of time-evolution based methods highly non-trivial. Since it is essential to work with infinite systems, approaches that treat the environment perturbatively might not be able to capture the correct long-time behavior. Due to these difficulties, many theoretical approaches are restricted to small interacting regions, weak coupling to reservoirs, or translationally-invariant systems. 

In this paper, we present a method to approximately determine the stationary-state of open, interacting quantum wires. The system is driven out of equilibrium via a coupling to non-interacting reservoirs that initially feature different chemical potentials (i.e., a bias voltage). To be precise, we employ the so-called functional renormalization group (FRG)\cite{Metzner2012} in a Keldysh-contour formulation.\cite{Jakobs2007,Gezzi2007,Karrasch2010,Kennes2012} The FRG treats the two-particle interaction in a perturbative sense -- although it still includes an infinite resummation of terms of arbitrary order -- but accounts for the reservoirs exactly. In contrast to previous approaches to this problem,\cite{Jakobs2007} we incorporate second-order contributions and can therefore describe inelastic processes and heating effects. Due to the significant numerical cost, second-order FRG schemes have so far only been implemented for electronic systems in thermal equilibrium\cite{Karrasch2008,Heyder2014,Sbierski2017,Markhof2018,Weidinger2019} as well as for the single impurity Anderson model out of equilibrium.\cite{Jakobs2010b}

In this paper, we close this gap and implement a second-order Keldysh FRG approximation (using a reservoir cutoff) for quantum wires. Solving the corresponding flow equations is highly demanding and requires advanced numerical techniques. By combining a semi-analytic solution with MPI parallelization, we can treat systems of up to 60 interacting lattices sites. 

It is fair to say that there is no `gold standard' available to study the non-equilibrium steady-state of an interacting, open, one-dimensional quantum system. While the density-matrix renormalization group \cite{schmitteckert04,schmitteckert10,fhm1,fhm2} as well as the numerical renormalization group\cite{nrg1,nrg2} are considered to be numerically-exact, they are restricted to short time scales or short chains.
Iterative path integral or quantum Monte Carlo based approaches are generically restricted to small-to-intermediate two-particle interactions strength or short times, respectively.\cite{iterp,PhysRevB.82.205323,qmc1,qmc2,qmc3} Exact, Bethe-ansatz based methods have been developed to study the out-of-equilibrium steady state,\cite{Bertini2016,Castro-Alvaredo2016} but these powerful tools are limited to closed, integrable systems. Other approaches such as continuous unitary transformations\cite{floweq1} struggle to capture the emergent, collective behavior of one-dimensional systems. For a further discussion, we refer a recent review article on finite-temperature transport in one-dimensional systems.\cite{bertini2020finitetemperature}

We structure this exposition as follows. We first introduce the general model Hamiltonian (Sec.~\ref{sec:class_of_models}) and give a brief overview of Keldysh Green's functions (Sec.~\ref{sec:greenfunc}). The novel, second-order FRG scheme is discussed in Sec.~\ref{sec:sfrgFinite}; we put a particular emphasis on how to solve the flow equations in an efficient, highly-parallelized way. Results are presented in Sec.~\ref{sec:tb_chains}. First, our method is benchmarked in the equilibrium limit, which is well-understood (Sec.~\ref{ssec:eq_results}). Non-equilibrium is discussed in Sec.~\ref{ssec:bias_results}. We demonstrate that the FRG data depends strongly on the choice of the cutoff scheme. This is particularly severe out-of-equilibrium where a) no physical arguments exists in favor of a certain cutoff, and b) secular higher-order terms appear, which are only partly included in a our approach. In a nutshell, a straightforward second-order, reservoir-cutoff FRG framework is highly-demanding yet inadequate to study interacting quantum wires out of equilibrium.


\section{Class of models discussed}
\label{sec:class_of_models}
In this paper, we consider time-independent, fermionic models with a finite number \(N\in\mathbb{N}\) of interacting degrees of freedom:
\begin{equation}\label{eq:ham_gen}
    H_\mathrm{chain}=\sum_{i,j=1}^N h_{ij} c_i^\dagger c_j +\frac{1}{4}\sum_{i,j,k,l=1}^N v_{ijkl} c_i^\dagger c_j^\dagger c_l c_k.
\end{equation}
Later on, we will discuss the concrete example of a tight-binding chain. With this application in mind, we refer to Eq.~(\ref{eq:ham_gen}) as the \emph{chain}.
In order to devise a numerically-efficient FRG scheme, it will be crucial that \(v\) is short-ranged. An example, and the focus of this paper, is a nearest-neighbor interaction. Importantly, we impose no restrictions on \(h\). The chain is quadratically coupled to a finite number \(N_\mathrm{res}\) of infinite, non-interacting reservoirs:
\begin{equation}
    \begin{split}
    H^\nu_\mathrm{res}&=\sum_k \epsilon^\nu_k a_{k,\nu}^\dagger a_{k,\nu}^\vdag,\\
    H^\nu_\mathrm{coup}&=\sum_{i,k} t^\nu_{i,k} c_i^\dag a_{k,\nu}^\vdag\ +\ \mathrm{h.c.},\\
    H_\mathrm{tot}&=H_\mathrm{chain}+\sum_{\nu=1}^{N_\mathrm{res}} H^\nu_\mathrm{res}+H^\nu_\mathrm{coup}.
    \end{split}
\end{equation}
For the discussion of the computational complexity of our scheme, we assume \(N_\mathrm{res}\ll N\).

The system is initially prepared in a product state of an arbitrary quadratic state within the chain and thermal equilibrium of the (decoupled) reservoirs; the latter is fully characterized by temperatures $T_\nu$ and chemical potentials \(\mu_\nu\). The influence of a reservoir on the chain can be described by the following retarded and Keldysh hybridization functions:
\begin{equation}\label{eq:self_hybrid}\begin{split}
    \Gamma^\tn{ret/K}(\omega)&=\sum_{\nu=1}^{N_\tn{res}}\Gamma^{\nu,\tn{ret/K}}(\omega),\\
    \Gamma^{\nu,\tn{ret}}_{ij}(\omega) & = \sum_k t^\nu_{i,k}t^{\nu*}_{j,k} \frac{1}{\omega-\epsilon^\nu_k+\I 0^+},\\
    \Gamma^{\nu,\tn{K}}_{ij}(\omega)&= [1-2n^\nu(\omega)]\, 2\I\, \textnormal{Im }\Gamma^{\nu,\mathrm{ret}}_{ij}(\omega),
\end{split}\end{equation}
where $n^\nu(\omega)$ is the Fermi function:
\begin{equation}
n^\nu(\omega) = \frac{1}{1+\exp[(\omega-\mu_\nu)/T_\nu]}.
\end{equation}
We assume that every part of the chain features a decay channel into at least one of the reservoirs. This is essential in order to obtain a well-defined steady state which is independent of the initial preparation of the chain.

In this paper, we exclusively work with reservoirs that feature a flat density of states; this so-called wide-band limit is justified if the bandwidth of the reservoirs exceeds all other energy scales. Moreover, we assume that the reservoirs are either at zero or infinite temperature. Eq.~(\ref{eq:self_hybrid}) then takes the simpler form
\begin{equation}\begin{split}
    \label{eq:restrRes}
    \Gamma^{\nu,\mathrm{ret}}(\omega)& =-\I \Gamma^\nu,\\
    \Gamma^{\nu,\mathrm{K}}(\omega)& =-2\I[1-2n^\nu(\omega)]\Gamma^\nu\\& =  -2\I\begin{cases}\sgn(\omega-\mu_\nu)\Gamma^\nu & T_\nu=0 \\ 0 & T_\nu=\infty,\end{cases}
\end{split}\end{equation}
where \(\Gamma^\nu\in\mathbb{C}^{N\times N}\) are positive, hermitian matrices characterizing the coupling to the individual reservoirs. Note that all infinite-temperature reservoirs do not contribute to the Keldysh component.

\section{Green's functions}
\label{sec:greenfunc}

Out of equilibrium, the natural language to describe correlation functions is
the Keldysh formalism. We assume familiarity and refer the reader to other works for a thorough
introduction.\cite{Rammer1986} To make this paper self-contained, however, we will briefly introduce our notation and recapitulate some key concepts.

The single-particle Green's functions in the stationary state take the form
\begin{equation}
    G(\omega)=\begin{pmatrix}
        G^{11}(\omega) & G^{12}(\omega)\\
        G^{21}(\omega) & G^{22}(\omega)
    \end{pmatrix}=\begin{pmatrix}
        G^\mathrm{ret}(\omega) & G^\mathrm{K}(\omega)\\
        0 & G^\mathrm{adv}(\omega)
    \end{pmatrix}.
\end{equation}
The retarded component reads
\begin{equation}\label{eq:gr_from_self}
    \begin{split}
        G^\mathrm{ret}_{ij}(t,t')&=G^\mathrm{ret}_{ij}(t-t')=-\I\theta(t-t')\left\langle \left[c_j^\dagger(t'),c_i(t)\right]_+\right\rangle,\\
    G^\mathrm{ret}_{ij}(\omega)&=\int_{-\infty}^\infty \mathrm{d}t \mathrm{e}^{\I\omega t} G^\mathrm{ret}_{ij}(t)
    =G^\mathrm{adv}_{ji}(\omega)^*,
    \end{split}
\end{equation}
and is related to the non-interacting retarded Green's function $g^\tn{ret}(\omega)$ via the Dyson equation:
\begin{equation}\label{eq:dysonret}\begin{split}
        G^\mathrm{ret}(\omega) & =\frac{1}{g^\mathrm{ret}(\omega)^{-1}-\Sigma^\mathrm{ret}(\omega)}, \\
        g^\mathrm{ret}(\omega)&=\frac{1}{\omega-h- \Gamma^{\mathrm{ret}}(\omega)},
\end{split}\end{equation}
where the self-energy $\Sigma^\tn{ret}$ is associated with the two-particle interaction $v_{ijkl}$. The Keldysh component is given by
\begin{equation}
    \begin{split}
        G^\mathrm{K}_{ij}(t-t')&=\I\left[ \left\langle c_{j}^\dagger(t') c_i(t)\right\rangle-\left\langle c_i(t) c_{j}^\dagger(t') \right\rangle\right],\\
    G^\mathrm{K}(\omega)&=\int_{-\infty}^\infty \mathrm{d} t \mathrm{e}^{i\omega t} G^\mathrm{K}(t),
\end{split}
\end{equation}
and the corresponding Dyson equation takes the form
\begin{equation}\label{eq:gk_from_self}
    \begin{split}
         G^\mathrm{K}& = G^\mathrm{ret}[ (g^\mathrm{ret})^{-1} g^\mathrm{K}(g^\mathrm{adv})^{-1}+  \Sigma^\mathrm{K}] G^\mathrm{adv}\\
         & = G^\mathrm{ret}\Big[\Gamma^{\mathrm{K}} +  \Sigma^\mathrm{K}\Big] G^\mathrm{adv},
    \end{split}
\end{equation}
where we have exploited that
\begin{equation}\label{eq:gk}
g^\mathrm{K}=g^\mathrm{ret}\Gamma^{\mathrm{K}} g^\mathrm{adv}. 
\end{equation}

All quantities in Eqs.~(\ref{eq:gr_from_self}) and (\ref{eq:gk_from_self}) are matrices defined by two single-particle indices as well as a single frequency. To simplify the notation, we will frequently employ multi-indices \(1=(i_1, \alpha_1)\) that include both this single-particle index $i_1$ as well as the Keldysh index $\alpha_1\in\{1,2\}$. The frequency-dependence will be denoted separately.

If the entire system is in an equilibrium configuration ($T_\nu=T$, $\mu_\nu=\mu$), the Green's functions fulfill the fluctuation-dissipation theorem (FDT):
\begin{equation}\label{eq:fluc_dis}
    G^\mathrm{K}(\omega)=\left[1-2n(\omega)\right]\left[ G^\mathrm{ret}(\omega)-G^\mathrm{adv}(\omega) \right].
\end{equation}
Out of equilibrium, this no longer holds true, but the Keldysh Green's function
can always be expressed via an effective distribution function
\(n^\mathrm{eff}(\omega)\in\mathbb{C}^{N\times N}\):
\begin{equation}\label{eq:fluc_dis_eff}
    G^\mathrm{K}(\omega)=
        G^\mathrm{ret}(\omega)\left[1-2n^\mathrm{eff}(\omega) \right]
        -\left[1-2n^\mathrm{eff}(\omega) \right]G^\mathrm{adv}(\omega) .
  \end{equation}
At constant \(\omega\), this is a \emph{Sylvester equation}, which can be solved
if \(G^\mathrm{ret}(\omega)\) and \(G^\mathrm{adv}(\omega)\) have no common eigenvalues.
In equilibrium, the distribution function \(n^\mathrm{eff}(\omega)= n(\omega)\mathbbm{1}\) becomes diagonal and one recovers the fluctuation-dissipation theorem. Out of equilibrium, \(n^\mathrm{eff}(\omega)\) provides an intuitive extension of the equilibrium distribution function.

\section{Second order fRG formulation}
\label{sec:sfrgFinite}
The functional renormalization group is an implementation of the RG idea on the level of single-particle correlation functions.
One starts by introducing a low-energy cutoff $\la$ into the non-interacting Green's function, $g\to g^\la$. By virtue of this replacement, all vertex functions (such as the self-energy) acquire a $\la$-dependence; taking the derivative w.r.t.~\(\Lambda\) yields an infinite hierarchy of coupled differential (flow) equations that describe the changes of the vertex functions when the cutoff scale is altered. The flow equations are arranged in powers of the interaction strength $v_{ijkl}$. If one truncates them at a given order, one obtains a controlled approximation while still including an infinite resummation of higher-order contributions. An introduction to this method can be found in Refs.~\onlinecite{Metzner2012,kopietzBook}.

While the systems described in Sec.~\ref{sec:class_of_models} can easily be treated in a first-order scheme, such an approximation only produces frequency-independent corrections to the retarded self-energy.\cite{Jakobs2007,Gezzi2007,Karrasch2010,Kennes2012} Contributions to the Keldysh component of the self-energy are, however, expected to be essential in order to describe heating. Such effects may fundamentally change the phenomenology, especially in systems that are only weakly coupled to the environment. In this work, we aim to account for all second order terms. This has so far only been achieved in thermal equilibrium\cite{Karrasch2008,Heyder2014,Sbierski2017,Markhof2018,Weidinger2019} as well as for the single impurity Anderson model out of equilibrium.\cite{Jakobs2010b}

As a guide to the reader, we will now summarize the main characteristics of our second-order Keldysh FRG scheme. Auxiliary wide-band reservoirs are attached to all sites of the chain and serve as the cutoff.\cite{Karrasch2010,Jakobs2010a} The flow of the three-particle vertex is neglected. The key approximation of our approach is to modify the rhs of the flow equation for the two-particle vertex $\gamma^\la$ by dropping both its own feedback (i.e., replacing $\gamma^\la$ by the initial, bare interaction) as well as the feedback of the self-energy. The solution to the flow equation for $\gamma^\la$ is then nothing but second-order perturbation-theory in the presence of the additional reservoirs (i.e., at the scale $\la$). In contrast, the self-energy flow equation is solved in full and is not further approximated. 

Our FRG scheme is correct to second order in the interaction but still contains an infinite number of higher-order terms. Despite the seemingly crude approximation to the (vertex) flow equations, their solution requires elaborate numerical techniques. By performing the calculation on several hundreds of computing nodes via MPI parallelization, we can access systems of up to 60 lattice sites.

\subsection{Choice of the cutoff}
\label{ssec:cutoff_finite}
When choosing the cutoff, we have two main goals: (i) after truncation, we want to preserve as many symmetries as possible while (ii) aiming for a numerically efficient algorithm. It is not straightforward to achieve both of these goals simultaneously; hence, our approach is a compromise, and we cannot rule out that a different cutoff might yield different, and potentially better, results.

The physical system that we want to study is only weakly coupled to reservoirs. This results in a sharply peaked density of states, which poses a significant numerical problem. Thus, it is advantageous to employ a cutoff which introduces additional scattering. During the flow, physical decay processes are expected to be generated, which should guarantee sufficient smoothing as the cutoff scale is successively lowered. In order to preserve the fluctuation-dissipation theorem in the equilibrium limit (artificially breaking the FDT would lead, e.g., to anomalous heating), we refrain from employing a cutoff which modifies the distribution function. 

For these reasons, we use a reservoir cutoff scheme where additional, auxiliary wide-band reservoirs are attached to all sites of the chain. \cite{Karrasch2010,Jakobs2010a} These reservoirs are characterized by a hybridization \(\Lambda\), and their initial state is an equilibrium one governed by a temperature $T_\tn{cut}$ as well as a chemical potential $\mu_\tn{cut}$. In this paper, we will exclusively use $T_\tn{cut}=0$ or $T_\tn{cut}=\infty$. To be precise, we replace $\Gamma^\tn{ret}\to\Gamma^{\tn{ret},\la}$ as well as $\Gamma^\tn{K}\to\Gamma^{\tn{K},\la}$ and employ Eq.~(\ref{eq:restrRes}) to obtain
\begin{equation}\begin{split}
    \label{eq:restrResLa}
    \Gamma^{\mathrm{ret},\la}(\omega)& =\Gamma^\tn{ret}(\omega)-\I\la\mathbbm{1},\\
    \Gamma^{\mathrm{K},\la}(\omega) & =\Gamma^\tn{K}(\omega) -2\I [1-2n^\tn{cut}(\omega)]\la\mathbbm{1}.
\end{split}\end{equation}
The contribution of the physical reservoirs is given by Eq.~(\ref{eq:restrRes}). The low-frequency properties of the system are suppressed via Eq.~(\ref{eq:restrResLa}), which thus acts as an infrared cutoff. When the auxiliary reservoirs are decoupled ($\la=0$), the original, physical system is recovered. This cutoff has the advantage that the Green's functions at finite flow parameter have the same form as physical Green's functions. This allows us to simplify the flow equations significantly and moreover guarantees that causality is conserved automatically.\cite{Jakobs2010a} The same holds true for the fluctuation-dissipation theorem in the equilibrium limit if the temperature and chemical potential of the auxiliary reservoirs are chosen identical to the physical ones, $T_\nu=T_\tn{cut}=T$, $\mu_\nu=\mu_\tn{cut}=\mu$.

\subsection{Self-energy flow equation}

The flow of the self-energy is given by\cite{Metzner2012,kopietzBook}
\begin{equation}
    \label{eq:fin_fo_flow}
    \begin{split}
    &\partial_\Lambda \Sigma_{1'1}^\la(\omega)
    =-\frac{\I}{2\pi} \int d\Omega \sum_{22'}\gamma^\la_{1'2'12}(\omega, \Omega,\omega, \Omega) S^\la_{22'}(\Omega),
\end{split}
\end{equation}
where \(\gamma^\la\) denotes the one-particle irreducible two-particle vertex function, and $S^\la$ is the single-scale propagator:
\begin{equation}\label{eq:single_scale}
S^\Lambda(\omega) =-G^\Lambda(\omega) \{\partial_\Lambda [g^{\Lambda}(\omega)^{-1}]\}G^\Lambda(\omega) = \partial_\Lambda^* G^\Lambda(\omega).
\end{equation}
Here, $\partial_\Lambda^*$ indicates a derivative that acts only on the explicit $\Lambda$-dependence of the cutoff (but not on $\Sigma^\Lambda$). As a reminder, we note that the multi-indices \(1',2',\dots\) contain the single-particle as well as the Keldysh indices. The retarded part of $S^\la$ takes the form
\begin{equation}
S^{\tn{ret},\la} = \I G^{\tn{ret},\la}G^{\tn{ret},\la},
\end{equation}
where we have used Eqs.~(\ref{eq:dysonret}) and (\ref{eq:restrResLa}). The Keldysh component is given by
\begin{equation}\begin{split}
S^{\tn{K},\la}  = &\,S^{\tn{ret},\la}\Big[\Gamma^{\tn{K},\la}+\Sigma^{\tn{K},\la}\Big] G^{\tn{adv},\la} \\
+ &\, G^{\tn{ret},\la}\Big[\Gamma^{\tn{K},\la}+\Sigma^{\tn{K},\la}\Big] S^{\tn{adv},\la} \\
-& 2\I [1-2n^\tn{cut}(\omega)]G^{\tn{ret},\la} G^{\tn{adv},\la},
\end{split}\end{equation}
where we have employed Eqs.~(\ref{eq:gk_from_self}) and (\ref{eq:restrResLa}).

\subsection{Vertex flow equation}

Our goal is to extend the functional renormalization group beyond leading order in a way that includes all second order contributions but that is still numerically feasible. This will allow us to analyze the effect of inelastic processes and to understand how they modify the first-order behavior. 

Time-translational invariance enforces energy conservation, and we can parametrize the frequency-dependence of the two-particle vertex \(\gamma^\la(\omega_{1'},\omega_{2'},\omega_1,\omega_2)=\gamma^\la(\Pi,X,\Delta)\) via the variables:
\begin{equation}
    \label{eq:freq_trafo}
    \begin{split}
        \Pi& =\omega_1+\omega_2=\omega_{1'}+\omega_{2'},\\
        X& =\omega_{2'}-\omega_1=\omega_{2}-\omega_{1'},\\
        \Delta &=\omega_{1'}-\omega_1=\omega_{2}-\omega_{2'}.
    \end{split}
\end{equation}
The flow-equation for the two-particle vertex function then reads:
\begin{widetext}
\begin{equation}
    \label{eq:sord}
    \begin{split}
        \partial_\Lambda \gamma^\Lambda_{1'2'12}(\Pi,X,\Delta)=&\frac{\I}{2\pi}\int d\Omega\sum_{33'44'}\\
        &\hspace{-2cm}\gamma^\Lambda_{1'2'34}\left(\Pi, \Omega+\frac{X-\Delta}{2}, \Omega-\frac{X-\Delta}{2}\right)S^\Lambda_{33'}\left(\frac{\Pi}{2}-\Omega\right)G^\Lambda_{44'}\left(\frac{\Pi}{2}+\Omega\right)\gamma^\Lambda_{3'4'12}\left(\Pi,\frac{X+\Delta}{2}+\Omega, \frac{X+\Delta}{2}-\Omega\right)\\
       &\hspace{-2.4cm}+\gamma^\Lambda_{1'4'32}\left(\frac{\Pi+\Delta}{2}+\Omega, X, \frac{\Pi+\Delta}{2}-\Omega\right)\biggl[S^\Lambda_{33'}\left(\Omega-\frac{X}{2}\right)G^\Lambda_{44'}\left(\Omega+\frac{X}{2}\right)+\\
        &\hspace{3.86cm}G^\Lambda_{33'}\left(\Omega-\frac{X}{2}\right)S^\Lambda_{44'}\left(\Omega+\frac{X}{2}\right)\biggr]\gamma^\Lambda_{3'2'14}\left(\Omega+\frac{\Pi-\Delta}{2}, X, \Omega- \frac{\Pi-\Delta}{2}\right)\\
        &\hspace{-2.4cm}-\gamma^\Lambda_{1'3'14}\left(\Omega+\frac{\Pi-X}{2}, \Omega-\frac{\Pi-X}{2}, \Delta\right)\biggl[S^\Lambda_{33'}\left(\Omega-\frac{\Delta}{2}\right)G^\Lambda_{44'}\left(\Omega+\frac{\Delta}{2}\right)+\\
        &\hspace{3.91cm}G^\Lambda_{33'}\left(\Omega-\frac{\Delta}{2}\right)S^\Lambda_{44'}\left(\Omega+\frac{\Delta}{2}\right)\biggr]\gamma^\Lambda_{4'2'32}\left(\frac{\Pi+X}{2}+\Omega, \frac{\Pi+X}{2}-\Omega,\Delta\right)\\
        &+\mathcal{O}(U^3),
    \end{split}
\end{equation}
\end{widetext}
where we already truncated the otherwise infinite hierarchy of differential equations by neglecting the flow of the three-particle vertex. This approximation is controlled in a perturbative sense and all terms neglected are at least of third order in the interaction $v_{ijkl}$, which we symbolically denote as \(\mathcal{O}\left(U^3\right)\).

If the frequency space is discretized via a grid with \(N_\Omega\) points, the resulting two-particle vertex has $\ord{N^4 N_\Omega^3}$ non-vanishing entries. Computing the rhs of an individual element is associated with a cost of \(\ord{N^4N_\Omega}\),\footnote{In this argument, we assume the grid is chose fine enough to approximate the integration.} resulting in an complexity class of
\begin{equation}
    \text{full 2nd order fRG}\in \ord{N^8 N_\Omega^4}
\end{equation}
to compute the flow of the two-particle vertex (the flow of the self-energy is significantly cheaper computationally). This is impractical even for small systems, and further approximations need to be devised. This will be the subject of Sec.~\ref{eq:sec_vertexsimple}.

\subsection{Initial condition}

When the coupling to the reservoirs is large ($\Lambda\to\infty$), the vertex functions can be obtained analytically:
\begin{equation}
    \label{eq:ini_cond}
    \begin{split}
        \Sigma^{\mathrm{ret}, \Lambda\to\infty}_{i'i}&=\frac{1}{2}\sum_j v_{i'jij},~
        \Sigma^{\mathrm{K},\Lambda\to\infty}_{i'i}=0,~
        \gamma^{\Lambda\to\infty}_{1'2'12}=\bar v_{1'2'12},
    \end{split}
\end{equation}
where we introduced the Keldysh-space version of the two-particle interaction
\begin{equation}
    \bar v_{1'2'12}=\begin{cases} \frac{1}{2} v_{i_{1'} i_{2'} i_1 i_2} & \alpha_{1'}+\alpha_{2'}+\alpha_1+\alpha_2\ \text{odd}\\
            0 & \text{otherwise.}\end{cases}
\end{equation}
The initial value of the retarded self-energy is frequency independent and can therefore be absorbed into the non-interacting Hamiltonian $h$.

\subsection{Simplification of the vertex flow equation}
\label{eq:sec_vertexsimple}

As a first step to reduce the complexity of the vertex flow equation (\ref{eq:sord}), we replace \(\gamma^\la\) with its initial value \(\bar v\) on the rhs (i.e., we remove the feedback of the two-particle vertex into its own flow equation). As \(\gamma^\la=\bar v+\ord{U^2}\), this only generates an error of \(\ord{U^3}\). The flow equation can then naturally be split up into three independent terms, \(\gamma^\la=\bar v+\gamma^{\text{p},\la}(\Pi)+\gamma^{\text{x},\la}(X)+\gamma^{\text{d},\la}(\Delta)\): 
\begin{equation}
    \label{eq:chan_decomp_flow}
    \begin{split}
        \partial_\Lambda \gamma^{\mathrm{p},\Lambda}_{1'2'12}(\Pi)=&\frac{\I}{2\pi}\int d\Omega\sum_{33'44'}\bar v_{1'2'34}S^\Lambda_{33'}G^\Lambda_{44'}\bar v_{3'4'12},\\
        \partial_\Lambda \gamma^{\mathrm{x},\Lambda}_{1'2'12}(X)=&\frac{\I}{2\pi}\int d\Omega\sum_{33'44'}\bar v_{1'4'32}\Big[S^\Lambda_{33'}G^\Lambda_{44'}\\ & \hspace*{2.95cm}+G^\Lambda_{33'}S^\Lambda_{44'}\Big]\bar v_{3'2'14},\\
        \partial_\Lambda \gamma^{\mathrm{d},\Lambda}_{1'2'12}(\Delta)=&\frac{-\I}{2\pi}\int d\Omega\sum_{33'44'}\bar v_{1'3'14}\Big[S^\Lambda_{33'}G^\Lambda_{44'}\\&\hspace*{2.95cm}+G^\Lambda_{33'}S^\Lambda_{44'}\Big]\bar v_{4'2'32},
    \end{split}
\end{equation}
with the initial condition being $\gamma^{\alpha,\la\to\infty}=0$, $\alpha=\tn{p,x,d}$. We have omitted the frequency arguments of $S^\la$ and $G^\la$; they are given explicitly in Eq.~(\ref{eq:sord}). The flow equation of the self-energy, which is not subject to any further approximations, can be decomposed correspondingly:
\begin{equation}\label{eq:flow_self_channel_ssfinite}
\begin{split}
    &\partial_\Lambda\Sigma^\Lambda_{1'1}(\omega) =-\frac{\I}{2\pi} \int d\Omega\sum_{22'}  S^\Lambda_{22'}(\Omega)\times\\
    &\Big[\bar v_{1'2'12} + \gamma^{\mathrm{p},\Lambda}_{1'2'12}(\Omega+\omega)+\gamma^{\mathrm{x},\Lambda}_{1'2'12}(\Omega-\omega)+\gamma^{\mathrm{d},\Lambda}_{1'2'12}(0)\Big].
\end{split}\end{equation}
Note that \(\gamma^{\mathrm{d},\la}(\Delta)\) is only needed at \(\Delta=0\).

Secondly, we remove the self-energy feedback on the rhs of the flow equation for the two-particle vertex by replacing \(G^\la\to g^\la\) as well as $S^\la\to s^\la= \partial_\la g^\la$, which again only introduces errors of \(\ord{U^3}\). At first sight, this might appear to make the problem more complicated from a numerical perspective as without feeding back the inelastic contributions of the self-energy, the Green's functions might be sharply peaked and evaluating the integrals might become more difficult. However, our approximation allows us to analytically integrate the flow equations. By exploiting that \(\partial_\Lambda g_{33'}^\la g_{44'}^\la= s_{33'}^\la g_{44'}^\la+g_{33'}^\la s_{44'}^\la\) as well as \(\bar v_{1'2'12}=-\bar v_{2'112}=-\bar v_{1'2'21}\) and by renaming $\Omega\to-\Omega$ in the case of $\gamma^{\mathrm{p},\Lambda}$ , we obtain
\begin{equation}
    \label{eq:chan_decomp_flow2}
    \begin{split}
        \gamma^{\mathrm{p},\Lambda}_{1'2'12}(\Pi)=&\frac{\I}{4\pi}\int d\Omega\sum_{33'44'}g^\Lambda_{33'}\left(\frac{\Pi}{2}-\Omega\right)g^\Lambda_{44'}\left(\frac{\Pi}{2}+\Omega\right)\\&\hspace*{3cm}\times \bar v_{1'2'34}\bar v_{3'4'12},\\[1ex]
        \gamma^{\mathrm{x},\Lambda}_{1'2'12}(X)=&\frac{\I}{2\pi}\int d\Omega\sum_{33'44'}g^\Lambda_{33'}\left(\Omega-\frac{X}{2}\right)g^\Lambda_{44'}\left(\Omega+\frac{X}{2}\right)\\& \hspace*{3cm}\times \bar v_{1'4'32}\bar v_{3'2'14},\\[1ex]
        \gamma^{\mathrm{d},\Lambda}_{1'2'12}(\Delta)=&\frac{-\I}{2\pi}\int d\Omega\sum_{33'44'}g^\Lambda_{33'}\left(\Omega-\frac{\Delta}{2}\right)g^\Lambda_{44'}\left(\Omega+\frac{\Delta}{2}\right)\\& \hspace*{3cm}\times \bar v_{1'3'14}\bar v_{4'2'32}.
    \end{split}
\end{equation}
Eq.~(\ref{eq:chan_decomp_flow2}) is nothing but the perturbation-theory result for the two-particle vertex in the presence of a finite flow parameter \(\Lambda\). The cutoff enters in the bare Green's functions $g^\la$.

While at \(T=0\) some components of the Green's functions and single-scale propagators are discontinuous, this is not true for the vertex functions, which one can understand as follows: The rhs of the flow equations (\ref{eq:chan_decomp_flow2}) is governed by a convolution of two functions $g^\Lambda$ that decay sufficiently quickly for large frequencies; this yields a continuous function.

The new flow equations (\ref{eq:chan_decomp_flow2}) each depend on a single frequency (and not on three frequencies), which drastically simplifies calculations. In addition, the dependence on the single-particle indices is reduced:
\begin{equation}
    \label{eq:sup_chan_two_part}
    \begin{split}
v_{i_{1'}i_{2'}\bullet\bullet}=0 \lor  v_{\bullet\bullet i_{1}i_{2}}=0~~&\Rightarrow~\gamma^{\text{p},\la}_{1'2'12}(\Pi)=0,\\
v_{i_{1'}\bullet\bullet i_{2}}=0 \lor  v_{\bullet i_{2'}i_{1}\bullet}=0~~&\Rightarrow~\gamma^{\text{x},\la}_{1'2'12}(X)=0,\\
v_{i_{1'}\bullet i_{1}\bullet}=0 \lor  v_{\bullet i_{2'} \bullet i_{2}}=0~~&\Rightarrow~\gamma^{\text{d},\la}_{1'2'12}(\Delta)=0.
    \end{split}
\end{equation}
This follows directly from Eq.~(\ref{eq:chan_decomp_flow2}); e.g., $\gamma^{\tn{p},\la}_{1'2'12}$ contains a term $v_{1'2'34}$ and thus vanishes for those indices $1',2'$ where $v_{1'2'34}=0$. For a nearest-neighbor interaction, Eq.~(\ref{eq:sup_chan_two_part}) simplifies to (other short-ranged interactions follow similarly)
\begin{equation}
    \begin{split}
        \gamma^{\text{p},\la}_{1'2'12}(\Pi   )&=0\hspace{0.5cm}  \forall\, |i_{1'}-i_{2'}|\neq 1 \lor |i_{1}-i_{2}|\neq 1,\\
        \gamma^{\text{x},\la}_{1'2'12}(X     )&=0\hspace{0.5cm}  \forall\, |i_{1'}-i_{2 }|   > 1  \lor |i_{2'}-i_{1}|>1,\\
        \gamma^{\text{d},\la}_{1'2'12}(\Delta)&=0\hspace{0.5cm}  \forall\, |i_{1'}-i_{1 }|   > 1 \lor |i_{2'}-i_{2}|>1.
    \end{split}\label{eq:sparseTwoPart}
\end{equation}
By virtue of Eqs.~(\ref{eq:sup_chan_two_part}) and (\ref{eq:sparseTwoPart}), only $\ord{N^2}$ terms are generated in each channel $\gamma^{\alpha,\la}$, $\alpha=\tn{p,x,d}$ of the vertex flow equation (\ref{eq:chan_decomp_flow2}). Moreover, the summation over the single-particle indices $i_3$, $i_{3'}$, $i_4$, and $i_{4'}$ in Eq.~(\ref{eq:chan_decomp_flow2}) only involves a limited number of terms and does not scale with $N$. The numerical cost of computing $\gamma^\la$ at a given $\la$ is then given by
\begin{equation}\label{eq:frg_scaling1}
\underbrace{\ord{N^2 N_\Omega}}_{\text{\#components}}\underbrace{\ord{N_\Omega}}_{\text{integration}}.
\end{equation}
This does not include the cost of computing the Green's functions $g^\la$, which is expected to scale like \(\ord{N^3N_\Omega}\) and which should be done beforehand. While Eqs.~(\ref{eq:chan_decomp_flow2}) can be solved numerically, it turns out to be more efficient to employ a semi-analytical solution, which we will discuss in the next section.

\subsection{Analytically computing the perturbative two-particle vertex}
\label{sec:fin_ala_vert}

We will now derive a semi-analytic way to determine the rhs of Eq.~(\ref{eq:chan_decomp_flow2}) at a given value of the flow parameter $\la$. To improve readability, we use the short hand notation for the effective (retarded) Hamiltonian
\begin{equation}
    \bar h=h+\Gamma^{\tn{ret},\la},
\end{equation}
where $\Gamma^{\tn{ret},\la}$ has been defined in Eq.~(\ref{eq:restrResLa}) and is frequency-independent for the case of physical wide-band reservoirs that we focus on exclusively [see Eq.~(\ref{eq:restrRes})]. As \(\bar h\) is not hermitian, it has separate left and right eigensystems:
\begin{equation}\label{eq:barheig}
    \begin{split}
        \bar h\left| q \right\rangle = \lambda_q  \left| q \right\rangle,~~~
        \left\langle \bar q \right| \bar h = \left\langle \bar q \right| \lambda_q.
    \end{split}
\end{equation}
The positivity of $\Gamma^{\tn{ret},\la}$ ensures that \(\Im(\lambda_q)<0\ \forall q\).
We can now rewrite the non-interacting retarded and advanced Green's functions as:
\begin{equation}
    \label{eq:grDecomp}
    \begin{split}
        g^{\tn{ret},\la}(\omega)&=\frac{1}{\omega-\bar h}=\sum_q \frac{1}{\omega-\lambda_q}\ket{q}\bra{\bar{q}}=\sum_q \frac{1}{\omega-\lambda_q}Q_q,\\
        g^{\tn{adv},\la}(\omega)&=\frac{1}{\omega-\bar h^\dagger}=\sum_q \frac{1}{\omega-\lambda^*_q}\ket{\bar q}\bra{q}=\sum_q \frac{1}{\omega-\lambda^*_q}Q^\dagger_q,
    \end{split}
\end{equation}
where we introduced the matrix \(Q_q=\ket{q}\bra{\bar q}\). Next, we simplify the Keldysh Green's function $g^{\tn{K},\la}(\omega)$. By virtue of Eq.~(\ref{eq:fluc_dis_eff}), $g^{\tn{K},\la}(\omega)$ can be expressed in terms of an effective distribution function $n^\tn{eff}(\omega)$, which can be related to the hybridization $\Gamma^{\tn{K},\la}$ via the Dyson equation (\ref{eq:gk}):
\begin{equation}\label{eq:sylvester2}\begin{split}
g^{\tn{K},\la}& =g^{\tn{ret},\la}\Gamma^{\tn{K},\la}g^{\tn{adv},\la}\\ &= 
g^{\tn{ret},\la}(1-2n^\tn{eff}) - (1-2n^\tn{eff})g^{\tn{adv},\la}\\[1ex]
&\Leftrightarrow \Gamma^{\tn{K},\la} = \bar h (1-2n^\tn{eff}) - (1-2n^\tn{eff})\bar h^\dagger.
\end{split}\end{equation}
All the reservoirs contribute additively to $n^\tn{eff}$, and the only frequency-dependence stems from the Fermi functions $n^\nu(\omega)$ and $n^\tn{cut}(\omega)$. The effective distribution function can thus be expressed in terms of frequency-independent operators $\eta_\nu$ and $\eta_\tn{cut}$:
\begin{equation}\begin{split}\label{eq:sylvesterIndiRes}
    1-2n^\tn{eff}(\omega) & = \sum_{\alpha=\nu,\tn{cut}} \eta_\alpha \left[1-2n^\alpha(\omega)\right]\\& = \sum_{\substack{\alpha=\nu,\tn{cut}\\ T_\alpha=0}}\eta_\alpha\sgn(\omega-\mu_\alpha),
\end{split}\end{equation}
where we have used that each of the reservoirs is held at either zero or infinite temperature. By comparing Eqs.~(\ref{eq:restrRes}), (\ref{eq:sylvester2}), and (\ref{eq:sylvesterIndiRes}), we obtain
\begin{equation}\begin{split}
    -2\I\Gamma^\nu&=\bar h \eta_\nu- \eta_\nu\bar h^\dagger,~~  -2\I\la\mathbbm{1}=\bar h \eta_\tn{cut}- \eta_\tn{cut}\bar h^\dagger.
\end{split}\end{equation}
The resulting equations are of a Sylvester form and can be solved via the Bartels-Stewart algorithm.\cite{Bartels1972} Note that a unique solution for $n^\tn{eff}$ exists if and only if \(\bar h\) has no real eigenvalues, which is equivalent to the statement that all degrees of freedom have a decay channel into one of the reservoirs.

Using Eqs.~(\ref{eq:grDecomp}) and (\ref{eq:sylvester2}), we can now express all the terms appearing on the rhs of Eq.~(\ref{eq:chan_decomp_flow2}) via complex-valued integrals. A specific example is given by
\begin{equation}
    \label{eq:exampleTwoGF}
    \begin{split}
    &\int \dOp \Omega\ g^{\tn{ret},\la}_{i_3i_{3'}}(\pm\Omega) g^{\tn{K},\la}_{i_4i_{4'}}(\Omega+\omega)\\
                                                                                  &=\int \dOp\Omega \sum_{q_1} \frac{1}{\pm\Omega-\lambda_{q_1}} \left(Q_{q_1}\right)_{i_3i_{3'}} \sum_{q_2}\sum_{\alpha}\sgn(\Omega+\omega-\mu_\alpha)\\
                                                                                   &\times\biggl[\frac{1}{\Omega+\omega-\lambda_{q_2}} \left( Q_{q_2}\eta_\alpha\right)_{i_4i_{4'}}-\frac{1}{\Omega+\omega-\lambda_{q_2}^*} \left(\eta_\alpha Q^\dag_{q_2}\right)_{i_4i_{4'}}\biggr]\\[1ex]
                                                                                   &=\pm\sum_{q_1q_2}\sum_{\alpha} \left(Q_{q_1}\otimes Q_{q_2} \eta_\alpha\right)_{i_3i_{3'}i_4i_{4'}}f_1(\pm \lambda_{q_1}, \lambda_{q_2}-\omega,\mu_\alpha)\\
                                                                                   &\hspace{1.5cm}-\left(Q_{q_1}\otimes \eta_\alpha Q^\dag_{q_2} \right)_{i_3i_{3'}i_4i_{4'}}f_1(\pm \lambda_{q_1}, \lambda_{q_2}^*-\omega,\mu_\alpha),
\end{split}
\end{equation}
where $\omega\in\{\Pi,X,\Delta\}$, we have shifted the integration variable $\Omega$, and introduced
\begin{equation}
    \begin{split}
    f_1(a,b,\mu)&=\int d\Omega \frac{1}{\Omega-a} \frac{1}{\Omega-b} \sgn(\Omega-\mu).
\end{split}
\end{equation}
Importantly, the frequency integrals $f_1$ in Eq.~(\ref{eq:exampleTwoGF}) do not depend on the single-particle indices $i_3,i_{3'},i_4,i_{4'}$ and can be computed analytically. A detailed account of this and of how to treat all other terms appearing on the rhs of Eq.~(\ref{eq:chan_decomp_flow2}) can be found in Appendix~\ref{ch:twoGF}.

The complexity of calculating the two-particle vertex for a given value of $\la$ using this strategy has a drastically different, and in some cases favorable, scaling:
\begin{equation}\label{eq:frg_scalingana}
\underbrace{\ord{N^2 N_\Omega}}_{\text{\#components}}~~\underbrace{\ord{N^2}}_{q_{1,2}\text{-summation}}.
\end{equation}
The main advantage compared to Eq.~(\ref{eq:frg_scaling1}) is the reduction from \(\ord{N_\Omega^2}\) to \(\ord{N_\Omega}\) by avoiding the numerical calculation of the integrals; this is achieved at the cost of an additional internal summation over \(N^2\) entries. Note that to obtain this complexity class, it is essential to first compute and store all \(Q_q, \eta_\alpha, Q_q\eta_\alpha, \eta_\alpha Q_q\) for a given $\la$.

\subsection{Frequency integrations}
\label{ssec:freg_integ}

\paragraph{Discretization of the self-energy}
We just illustrated that the frequency integration on the rhs of Eq.~(\ref{eq:chan_decomp_flow2}) can be carried out analytically. The integral that appears in the self-energy flow equation (\ref{eq:flow_self_channel_ssfinite}), however, needs to be performed numerically. It is thus necessary to discretize the frequency argument of $\Sigma^\la(\omega)$. In this work, we employ a fixed, equidistant frequency grid and assume that the self-energy is step-wise constant between grid points. We explicitly tested that our results are converged w.r.t.~to the grid parameters such as its number of elements $N_\Omega$, its spacing, and its largest frequency.

At the beginning (end) of the flow, the single-scale propagator as well as the two-particle vertex decay on a scale given by the coupling $\la$ to the auxiliary reservoir (by the physical bandwidth); hence, they need to be evaluated for large frequencies (for frequencies on the scale of the bandwidth).\footnote{Note that in order to evaluate observables at the end of the flow, the self-energy is required for all frequencies within the support of the Green's functions. For that reason, the largest frequency in our grid always has to be much larger than the physical bandwidth.} In order to avoid having to adapt the grid during the solution of the flow equations, we note that the first-order contribution to the self-energy is frequency independent and thus \(\Sigma_{1'1}^\la(\omega)-\Sigma_{1'1}^\la(\omega')\sim \ord{U^2}\). We can therefore always approximate the self-energy at arbitrarily large frequencies by its value at the largest frequency in our grid; this only leads to errors in \(\ord{U^3}\).

\paragraph{Indefinite integrals}
After discretizing the self-energy, the integral in Eq.~(\ref{eq:flow_self_channel_ssfinite}) can be performed numerically; we employ the \texttt{runge\_kutta\_cash\_karp54} implementation provided by \texttt{boost}.\cite{boost} The indefinite integral is recast as
\begin{equation}
    \int_{-\infty}^\infty \dOp \omega f(\omega)=\int_{-A}^A \dOp \omega f(\omega)+\int_{-\frac{1}{A}}^\frac{1}{A} \frac{\dOp \eta}{\eta^2}f\left(\frac{1}{\eta}\right),
    \label{eq:indefInteg}
\end{equation}
where we substituted \(\eta=\frac{1}{\omega}\) in the last term. Those terms in Eq.~(\ref{eq:flow_self_channel_ssfinite}) that involve a Keldysh Green's function feature discontinuities at every chemical potential $\mu_\nu$ and $\mu_\tn{cut}$ of any of the zero-temperature reservoirs. It is most efficient to split up the corresponding integrals such that none of them contains a discontinuity.

\paragraph{Lookup tables for vertex functions}
The two-particle vertex functions $\gamma^{\tn{p},\la}(\Omega+\omega)$ and $\gamma^{\tn{x},\la}(\Omega-\omega)$ that appear on rhs of the self-energy flow equation (\ref{eq:flow_self_channel_ssfinite}) need to be evaluated for $N_\Omega^2$ different frequency arguments in order to perform the convolution. Since the analytical approach outlined in Sec.~\ref{sec:fin_ala_vert} is numerically expensive, it is favorable to compute a lookup table for $\gamma^{\tn{p},\la}$ and $\gamma^{\tn{x},\la}$ on a given set of frequencies beforehand; it is efficient to (ab)use an integration routine to determine an optimal grid. In between grid points, the vertex functions are determined using a cubic spline interpolation. We made sure that our results are independent of the grid parameters. 

It is important to point out that the number of grid points needed to faithfully approximate the vertex functions does not necessarily scale with the system size $N$. We will come back to this issue in Sec.~\ref{sec:tb_chains}.

\subsection{Numerical Details, Parallelization}
\label{ssec:paralleli}

There are two reasons to devise a parallelized scheme to solve the FRG flow equations. First, the computational effort to treat large systems of \(\ord{50}\) sites is significant, and the wall-time performance can be enhanced greatly using parallelization. Secondly, the lookup table for the entire vertex function takes up large amounts of memory and cannot be stored on a single machine.\footnote{For a system of \(\ord{50}\) sites with a nearest-neighbor interaction, each $\gamma^{\alpha,\la}$ can be stored in a 1D-array of size \(\ord{10\mathrm{MB}}\) per frequency.} We aim at using MPI parallelization over hundreds of computing nodes, and it is essential to minimize the necessary communication between different machines.

The main idea is to calculate the rhs of $\partial_\la\Sigma^\la_{1'1}(\omega)$ in parallel by splitting up the indices $1',1$ into appropriate sets, which are handled by individual nodes. To be precise, we proceed as follows:
\begin{enumerate}
\item At a given $\la$, calculate the eigenvalues $\lambda_q$ and eigenvectors $\ket{q}$, $\bra{\bar q}$ of $\bar h$ [see Eq.~(\ref{eq:barheig})], determine $\eta_\alpha$ via Eqs.~(\ref{eq:sylvester2}) and (\ref{eq:sylvesterIndiRes}), and compute the products $\eta_\alpha Q_q$ and $Q_q\eta_\alpha$ $\forall q,\alpha$, $Q_q=\ket{q}\bra{\bar q}$.
\item Split up the indices $1',1$ into disjoint sets (see below). On each node, compute the rhs of the self-energy flow equation (\ref{eq:flow_self_channel_ssfinite}) for a given set and for all frequencies $\omega$. To this end, perform the following steps:
\item[3a.] Start a loop over the indices $2$ and $2'$ that appear on the rhs of Eq.~(\ref{eq:flow_self_channel_ssfinite}); these loops only contain a finite (small) number of terms $\gamma^{\tn{p},\la}_{1'2'12}$ and $\gamma^{\tn{x},\la}_{1'2'12}$ via Eqs.~(\ref{eq:sup_chan_two_part}) and (\ref{eq:sparseTwoPart}). 
\item[3b.] Perform the $\Omega$-integrals that involve $\gamma^{\tn{p/x},\la}_{1'2'12}(\Omega\pm\omega)$ numerically; $\gamma^{\tn{d},\la}_{1'12'2}$ is only needed at zero frequency. To this end, first calculate a frequency lookup table for $\gamma^{\tn{p/x},\la}_{1'2'12}(\tilde\omega)$. For a fixed $\tilde\omega$, proceed as follows:
 \item[3c.] Evaluate Eq.~(\ref{eq:chan_decomp_flow2}) in the spirit of Eq.~(\ref{eq:exampleTwoGF}); see also the appendix. In particular,\\[1ex]
 \hspace*{0.5cm} $\bullet$ loop over $q_{1,2}$ and $\alpha$,\\[1ex]
 \hspace*{1cm} $\bullet$  compute $f_{0,1,2}$ analytically,\\[1ex]
 \hspace*{1cm} $\bullet$ sum over $3,3',4,4'$, restrict single-particle\\\hspace*{1cm} $\phantom{\bullet}$ sums via Eqs.~(\ref{eq:sup_chan_two_part}) and (\ref{eq:sparseTwoPart}).
\end{enumerate}

The computational bottleneck of the FRG algorithm is the calculation of the two-particle vertex [see Eq.~(\ref{eq:frg_scalingana})], i.e., establishing the lookup table for $\gamma^{\tn{p/x},\la}_{1'2'12}(\tilde\omega)$ in step 3c. If the indices $1',1$ are grouped such that within one set the single-particle parts $i_{1'}$ and $i_1$ are chosen from the same spatial region, Eqs.~(\ref{eq:sup_chan_two_part}) and (\ref{eq:sparseTwoPart}) restrict $i_{2'}$ and $i_2$ to (roughly) the same region. Thus, each node needs to compute the lookup table for (roughly) disjoint sets of indices $1',2',1,2$, rendering parallelization highly-efficient. Note that while \(\gamma^{\mathrm{d},\la}_{1'2'12}\) is required for \(\ord{N}\) values of \(2',2\) on every machine, it only enters at zero frequency.

The only quantity that needs to be sent across nodes is the self-energy, which feeds back into its own flow equation (\ref{eq:flow_self_channel_ssfinite}). Note that the corresponding amount of data is small, and MPI parallelization is highly-efficient.

\begin{figure*}
\begin{tikzpicture}
\coordinate(A0) at (0.6cm,0cm);
\foreach \a in {1,2,...,7}{
    \if\a4%
        \node[circle, minimum size=0.6cm, right=1.1cm*\a of A0, thick](A\a){$\dots$};
    \else
        \node[draw, circle, minimum size=0.6cm, right=1.1cm*\a of A0, thick](A\a){};
    \fi;
}
\foreach[evaluate=\a as \an using int(\a+1)] \a in {1,2,...,6}
    \path[]
        (A\a.north east) edge [
          text=black,
          thick,
          shorten <=2pt,
          shorten >=2pt,
          bend left=50
          ] node[above]{\(t\)} (A\an.north west);
\foreach[evaluate=\a as \an using int(\a+1)] \a in {1,2,...,6}
    \path[]
        (A\a.south east) edge [
          text=black,
          thick,
          shorten <=2pt,
          shorten >=2pt,
          bend right=50,
          dashed
          ] node[below]{\(U\)} (A\an.south west);

      \coordinate[left=2cm of A1](r1);
      \draw[thick, shading=axis, left color=white, right color=custBlue] (r1)++(-90:1.5 and 0.5) arc(-90:90:1.5 and 0.5);
      \coordinate[right=2cm of A7](r2);
      \draw[thick, shading=axis, right color=white, left color=custBlue] (r2)++(270:1.5 and 0.5) arc(270:90:1.5 and 0.5);
    \path[]
        (r1)++(35:1.5 and 0.5) edge [
          text=black,
          thick,
          shorten <=2pt,
          shorten >=2pt,
          bend left=50
          ] node[above]{\(\Gamma\)} (A1.north west);
    \path[]
        (r2)++(145:1.5 and 0.5) edge [
          text=black,
          thick,
          shorten <=2pt,
          shorten >=2pt,
          bend right=50
          ] node[above]{\(\Gamma\)} (A7.north east);

\end{tikzpicture}
\caption{Pictorial representation of the system employed in Sec.~\ref{sec:tb_chains}. We consider a tight-binding chain ($N$ sites) of spinless fermions with a nearest-neighbor hopping $t$ and a nearest-neighbor interaction $U$. The chain is end-coupled to wide-band reservoirs. The reservoirs are characterized by a hybridization \(\Gamma\) and are initially in thermal equilibrium at zero temperature and with chemical potentials $\mu_\text{L,R}$. }
\label{fig:sys_finite_chain}
\end{figure*}
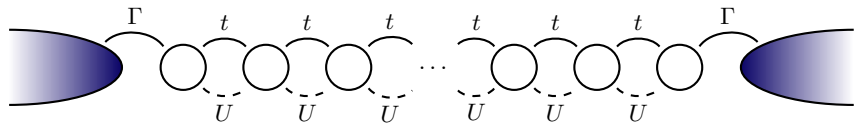%

\subsection{Perturbation theory}
\label{sec:pt}

Within the FRG scheme used in this work, the two-particle vertex is computed in second-order perturbation theory in $\bar v$. This is achieved by neglecting the feedback of both the self-energy and the two-particle vertex within the original flow equation (\ref{eq:sord}); the result is given in Eq.~(\ref{eq:chan_decomp_flow2}). As a reference, we now illustrate how to calculate the self-energy within perturbation theory in an easy fashion using the existing FRG formalism.

The first-order contribution to the self-energy can be obtained by replacing the single-scale propagator as well as the two-body vertex on the rhs of Eq.~\eqref{eq:flow_self_channel_ssfinite} by their lowest-order expansion (\(s^\la\) and \(\bar v\), respectively) and by integrating the resulting equation:\cite{christophdr}
\begin{equation}\label{eq:pt_first}
\begin{split}
    &\Sigma^{\tn{1PT},\la}_{1'1}(\omega) =-\frac{\I}{2\pi} \int d\Omega\sum_{22'}  g^\Lambda_{22'}(\Omega)\bar v_{1'2'12}.
\end{split}\end{equation}
In order to generalize this to second order, we compute the leading-order expansion of the single-scale propagator:
\begin{equation}\label{eq:pt_s}\begin{split}
S^\la&=\partial_\la^* G^\la =\partial_\la^*\left[g^\la+g^\la\Sigma^{\tn{1PT},\la}g^\la+\ord{U^2}\right]  \\&=s^\la+g^\la\Sigma^{\tn{1PT},\la} s^\la +s^\la\Sigma^{\tn{1PT},\la} g^\la+\ord{U^2} \\
&=s^\la +\ord{U}.
\end{split}\end{equation}
Since \(\gamma^{\mathrm{p/x/d},\la}\sim U^2\), the second-order contribution to the rhs of Eq.~\eqref{eq:flow_self_channel_ssfinite} that is associated with the x- and p-channel is given by
\begin{equation}\label{eq:flow_pt_px}
\begin{split}
    &-\frac{\I}{2\pi} \int d\Omega\sum_{22'}  s^\Lambda_{22'}(\Omega)\Big[\gamma^{\mathrm{p},\Lambda}_{1'2'12}(\Omega+\omega)+\gamma^{\mathrm{x},\Lambda}_{1'2'12}(\Omega-\omega)\Big]\\
    =&-\partial_\Lambda \frac{\I}{2\pi} \int d\Omega\sum_{22'}  g^\Lambda_{22'}(\Omega)\Big[\gamma^{\mathrm{p},\Lambda}_{1'2'12}(\Omega+\omega)\Big].
\end{split}\end{equation}
The derivative in the last line acts on $g^\la$ as well as on $\gamma^{\tn{p},\la}$, which yields the first and second term in the first line, respectively. The latter becomes clear if we use Eq.~(\ref{eq:chan_decomp_flow2}) and rename indices as well as the integration variables. Next, we discuss the second-order contributions to the rhs of Eq.~(\ref{eq:flow_self_channel_ssfinite}) that is attributed to the d-channel as well as to the single-scale propagator:
\begin{equation}\label{eq:flow_pt_d}
\begin{split}
    &-\frac{\I}{2\pi} \int d\Omega\sum_{22'} s^\Lambda_{22'}(\Omega)\gamma^{\mathrm{d},\Lambda}_{1'2'12}(0)\\
     &+\left\{s^\Lambda+g^\Lambda \Sigma^{\tn{1PT},\la} s^\Lambda+s^\Lambda \Sigma^{\tn{1PT},\la} g^\Lambda\right\}_{22'}(\Omega) \bar v_{1'2'12}\\
     =&-\partial_\Lambda\frac{\I}{2\pi} \int d\Omega\sum_{22'}  g^\Lambda_{22'}(\Omega)\Big[\bar v_{1'2'12} +\gamma^{\mathrm{d},\Lambda}_{1'2'12}(0)\Big],
\end{split}\end{equation}
where we have plugged in Eq.~(\ref{eq:pt_first}); the terms $g^\Lambda \Sigma^{\tn{1PT},\la} s^\Lambda\bar v$ can be identified with the term $g^\la \partial_\la\gamma^{\mathrm{d},\Lambda}$ via Eq.~(\ref{eq:chan_decomp_flow2}).

The second-order perturbation theory result for the self-energy can now be obtained by modifying the rhs of Eq.~(\ref{eq:flow_self_channel_ssfinite}) according to Eqs.~(\ref{eq:flow_pt_px}) and (\ref{eq:flow_pt_d}) and by integrating w.r.t.~$\la$:
\begin{equation}\label{eq:flow_pt}
\begin{split}
    &\Sigma^{\tn{2PT},\la}_{1'1}(\omega) =-\frac{\I}{2\pi} \int d\Omega\sum_{22'}  g^\Lambda_{22'}(\Omega)\times\\
    &\Big[\bar v_{1'2'12} + \gamma^{\mathrm{p},\Lambda}_{1'2'12}(\Omega+\omega)+\gamma^{\mathrm{d},\Lambda}_{1'2'12}(0)\Big].
\end{split}\end{equation}
We reiterate that \(\gamma^{\tn{p/d},\Lambda}\) denote the two-particle vertices within second-order perturbation theory [see Eq.~\eqref{eq:chan_decomp_flow2}]; \(\gamma^{\mathrm{x},\Lambda}\) does not appear separately. Note that Eq.~(\ref{eq:flow_pt}) is completely general and holds even if the first-order contribution $\Sigma^{\tn{1PT},\la}$ to the self-energy does not vanish.\cite{christophdr}

After plugging in the analytic expressions for \(\gamma^{\tn{p/d},\Lambda}\) derived in  Sec.~\ref{sec:fin_ala_vert}, the remaining frequency integral can be performed analytically using the techniques of Appendix~\ref{ch:twoGF}. While the ensuing expressions are lengthy, they are particularly helpful for large systems where the bare Green's functions are sharply-peaked and numerical integrations become demanding.

\section{Application to 1D chains}
\label{sec:tb_chains}
Here, we want to study one-dimensional metallic systems driven out of their equilibrium state. These systems are interesting even within equilibrium as a  Luttinger liquid state emerges at finite interactions, which replaces the non-interacting Fermi-liquid picture and signals the onset of critical, collective behavior.\cite{giamarchi,Mastropietro2013} The so-called Tomonaga-Luttinger model -- obtained by linearizing the dispersion of the electrons around the Fermi-points -- is the standard paradigm to understand the low-energy behavior of such systems.\cite{Tomonaga1950,Luttinger1963,Schoenhammer1997} However, motivating the Tomonaga-Luttinger model outside of the equilibrium (low-energy) realm becomes ambiguous. Recently, studies tried to extend its predictive power to  out-of-equilibrium setups by  taking into account non-linear contributions due to the band-curvature. This phenomenology was coined  non-linear Luttinger liquids, and interesting predictions with respect to non-equilibrium critical behavior have been made.\cite{Imambekov2009,Imambekov2012} However, due to the ambiguity of the low-energy assumption outside of equilibrium, these studies require a firm benchmark based on microscopic model calculations -- a goal that our novel FRG algorithm was set up to contribute to. 


\subsection{Model}

We consider a one-dimensional lattice of spinless fermions end-coupled to leads, which act as particle reservoirs.
We will restrict ourselves to the simplest case described by the Hamiltonian:
\begin{equation}
\begin{split}
        H_\mathrm{chain}&=H_\mathrm{tb}+H_\mathrm{int},\\
    H_\mathrm{tb}&=t\sum_{n=1}^{N-1} c_n^\dag c^\vdag_{n+1} +\mathrm{h.c.},\\
        H_\mathrm{int}&=U\sum_{n=1}^{N-1} \left(c_n^\dag c^\vdag_n -\frac{1}{2}\right)\left(c_{n+1}^\dag c^\vdag_{n+1} -\frac{1}{2}\right),
   \end{split}
\end{equation}
where \(N\) denotes the number of sites in the interacting chain, and $t$ and \(U\) are the strength of the nearest-neighbor hopping and interaction, respectively. Unless mentioned otherwise, we always set $t=1$.
Two wide-band reservoirs ($N_{\rm res}=2$) that we refer to as \emph{left} ($1=\mathrm{L}$) and \emph{right} ($2=\mathrm{R}$) are coupled to the ends of the chain and are characterized by the hybridization function
\begin{equation}\begin{split}
    \Gamma^{1,\mathrm{ret}}_{ij}&= -\I\Gamma^\mathrm{L}_{ij}=-\I\Gamma \delta_{i,1}\delta_{j,1},\\\Gamma^{2,\mathrm{ret}}_{ij}&=-\I\Gamma^\mathrm{R}_{ij}=-\I\Gamma\delta_{i,N}\delta_{j,N}.
\end{split}\end{equation}
Initially, the reservoirs are prepared in thermal equilibrium at zero temperature (\(T_\nu=0\)) and with chemical potentials \(\mu_1=\mu_\text{L}\), \(\mu_2=\mu_\text{R}\). A pictorial representation of the system is shown in Fig.~\ref{fig:sys_finite_chain}.

\begin{figure}[t]
    \includegraphics[width=\columnwidth]{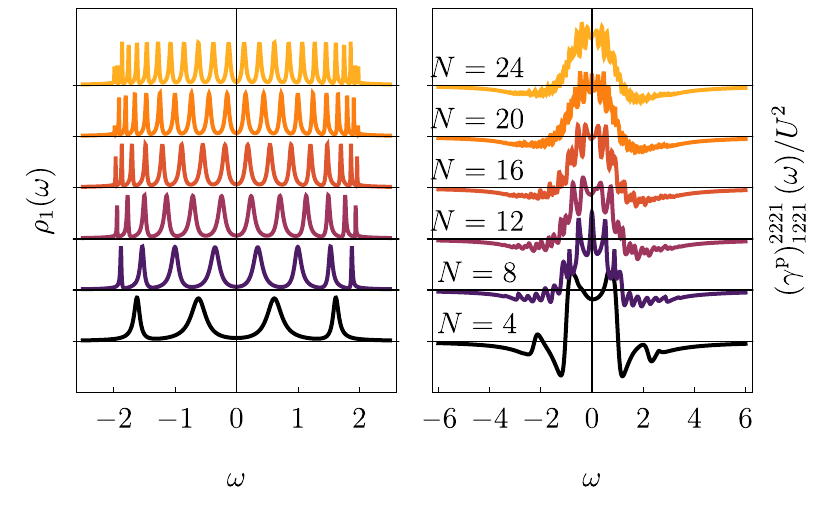}
    \caption{ \emph{Left panel}: The local density of states of a non-interacting chain ($U=0$) of length \(N=4,8,\dots,24\) (bottom to top) with \(\Gamma=0.2\) in thermal equilibrium ($\mu_\text{L}=\mu_\text{R}=0$). It features \(N\) peaks of width \(\sim \Gamma/N\). \emph{Right panel}: The \(\left(\gamma^\mathrm{p}\right)^{2221}_{1221}\)--component of the two-particle vertex (\ref{eq:chan_decomp_flow2}) at $\Lambda=0$ for \(\Gamma=0.2\) in thermal equilibrium ($\mu_\text{L}=\mu_\text{R}=0$). In contrast to the LDOS, the two-particle vertex features \(\ord{N^2}\) peaks and converges to a smooth function for $N\to\infty$.}
    \label{fig:smoothingTwoPartVert}
\end{figure}

Before we discuss the results, we briefly turn to a particularity specific to the model, which is relevant for the numerical efficiency of our algorithm.
In the absence of interactions, the system is perfectly coherent in the extended region between the reservoirs; the typical coherence time grows with \(N\). This is reflected in observables such as the local density of states (LDOS)
\begin{equation}
    \rho_i(\omega)=-\frac{1}{\pi}\Im[G^\mathrm{ret}_{ii}(\omega)],
\end{equation}
shown in Fig.~\ref{fig:smoothingTwoPartVert}. It features \(\ord{N}\) peaks of width \(\ord{\Gamma/N}\) and requires \(\ord{N}\) frequency points to be faithfully represented. This, however, does not imply that the number of discretization points $N_\Omega$ that appear within our algorithm needs to scale with \(N\), since the discretizations discussed in Sec.~\ref{ssec:freg_integ} are only used to represent the vertex functions, not the Green's functions themselves.
In contrast to the Green's functions, the two-particle vertex functions feature \(\ord{N^2}\) peaks of width \(\ord{\Gamma/N}\), making them smooth in the limit \(N\to \infty\). This is explicitly demonstrated in Fig.~\ref{fig:smoothingTwoPartVert}. A similar argument holds for the self-energy.

\subsection{Results in equilibrium}
\label{ssec:eq_results}

We will now test the second-order FRG approach in thermal equilibrium. Since this limit is well understood (in marked contrast to the non-equilibrium case), it provides a natural benchmark for our algorithm. In particular, we investigate the influence of the different choices of the temperature $T_\tn{cut}$ and chemical potential $\mu_\tn{cut}$ associated with the FRG cutoff, which yield different results due to the truncation of the flow equation hierarchy. Investigating the cutoff-dependence thus provides a way to study the reliability of our approximation scheme. We emphasize that the physical reservoirs are always held at zero temperature $T_\tn{L}=T_\tn{R}=0$.

In thermal equilibrium, it is natural to work with a cutoff whose temperature $T_\tn{cut}$ and chemical potential $\mu_\tn{cut}$ equals those of the physical reservoirs $\mu_\tn{L}=\mu_\tn{R}$, $T_\tn{L}=T_\tn{R}=0$. Out-of-equilibrium, however, it is a priori unclear what cutoff scheme one should employ; results are only meaningful if they do not depend on the particular choice of the cutoff.  We will now illustrate that the equilibrium data displays a strong cutoff-dependence. This is a first hint towards the inadequacy of our second-order FRG scheme in treating finite, interacting system in non-equilibrium.

\begin{figure}[t]
    \includegraphics[width=\columnwidth]{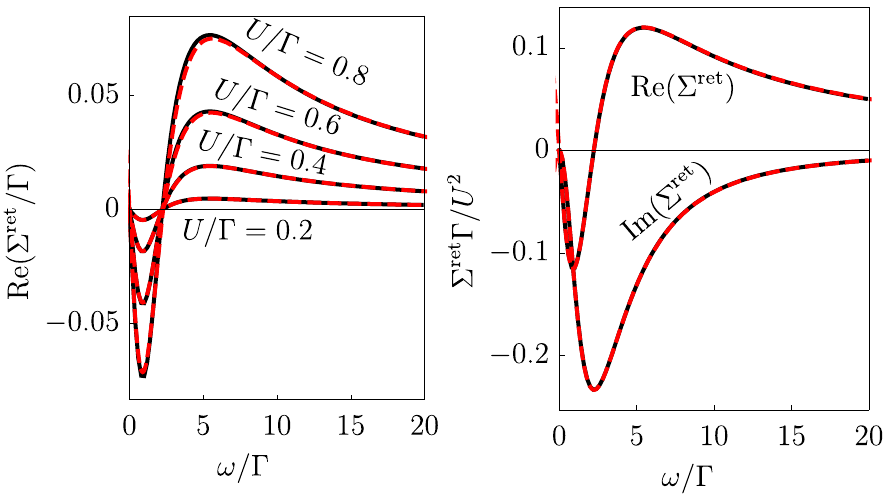}
    \caption{ \emph{Left panel}: Self-energy of the single impurity Anderson model in equilibrium ($N=2, t=0,\mu_\mathrm{L,R,cut}=0$). In the limit of small $U$, perturbation theory [solid lines, see Eq.~(\ref{eq:siam})] and FRG data (dashed lines) successively approach each other. \emph{Right panel}: If the FRG flow equations are modified according to Sec.~\ref{sec:pt}, perturbation theory is reproduced exactly.}
    \label{fig:siam}
\end{figure}%

\subsubsection{Single Impurity Anderson Model}

We first perform an instructive comparison to test the validity of our numerics. For a two-site system ($N=2$) with zero hopping ($t=0$) and $\mu_\tn{L}=\mu_\tn{R}=0$, one obtains a version of the single-impurity Anderson model. The self-energy can be computed in perturbation theory, and the result in equilibrium reads\cite{Hamiltonian1984}
\begin{equation}\label{eq:siam}\begin{split}
        \Sigma(&\omega)=U^2\int d\omega_1 d\omega_2 d\omega_3 \frac{\rho_0(\omega_1)\rho_0(\omega_2) \rho_0(\omega_3)}{\omega-\omega_1-\omega_2+\omega_3}\\ &\times\big[ n(\omega_1)n(\omega_2) n(-\omega_3) + n(-\omega_1)n(-\omega_2) n(\omega_3) \big],
\end{split}\end{equation}
where $n(\omega)=n^\tn{L}(\omega)=n^\tn{R}(\omega)$ is the Fermi function, and $\rho_0(\omega)$ denotes the non-interacting density of states at one of the sites. In the limit of small $U$, the FRG data successively approaches this result (see Fig.~\ref{fig:siam}, left panel). Moreover, one can exactly reproduce perturbation theory by modifying the flow equation according to Sec.~\ref{sec:pt}  (right panel of the figure).

\subsubsection{Fluctuation-dissipation theorem}

\begin{figure}[t]
    \includegraphics[width=\linewidth,clip]{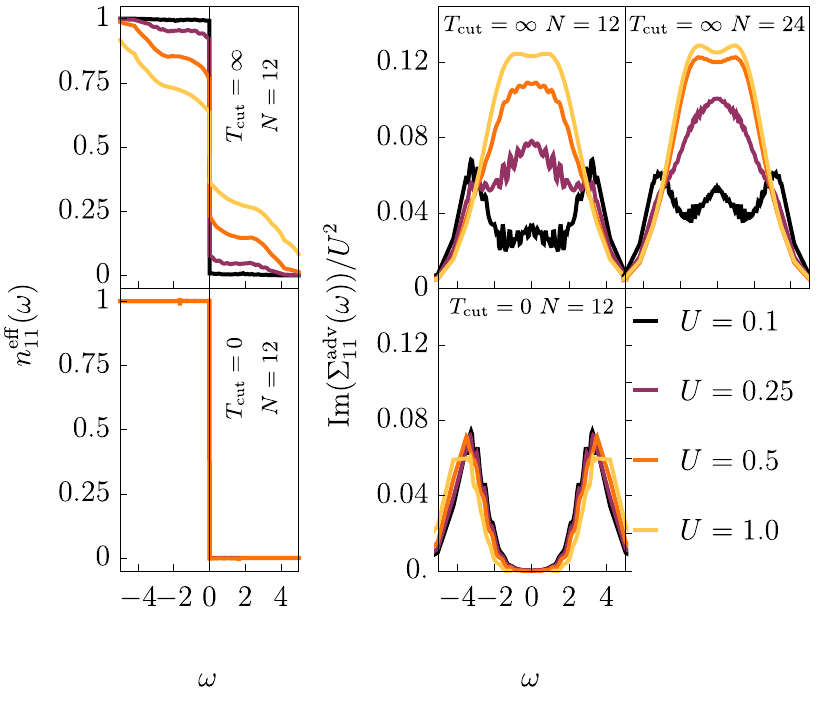}
    \caption{FRG results in equilibrium with \(\mu_\mathrm{L,R,cut}=0\), reservoir couplings \(\Gamma=0.2\), and various values of the interaction $U$.
    \emph{Left panel}: The effective distribution at the boundary of a system of \(N=12\) sites. The upper and lower panel show data obtained using an infinite- and zero-temperature cutoff scheme, respectively. Only the latter reproduces the correct equilibrium distribution function (\ref{eq:fluc_dis2}) stipulated by the fluctuation-dissipation theorem; the former yields artificial heating in form of a decreased discontinuity at the Fermi surface.    \emph{Right panel}: Imaginary part of the self-energy at the boundary, which serves as a measure for the magnitude inelastic processes at a given energy \(\omega\). In equilibrium, scattering at the Fermi surface is suppressed;\cite{giamarchi2003quantum,Samokhin1998} while data obtained with \(T_\mathrm{cut}=0\) reproduces this correctly, the infinite-temperature cutoff introduces unphysical scattering.
    }
    \label{fig:noBiasCutoffComp}
\end{figure}

\begin{figure}[t]
    \includegraphics[width=\columnwidth]{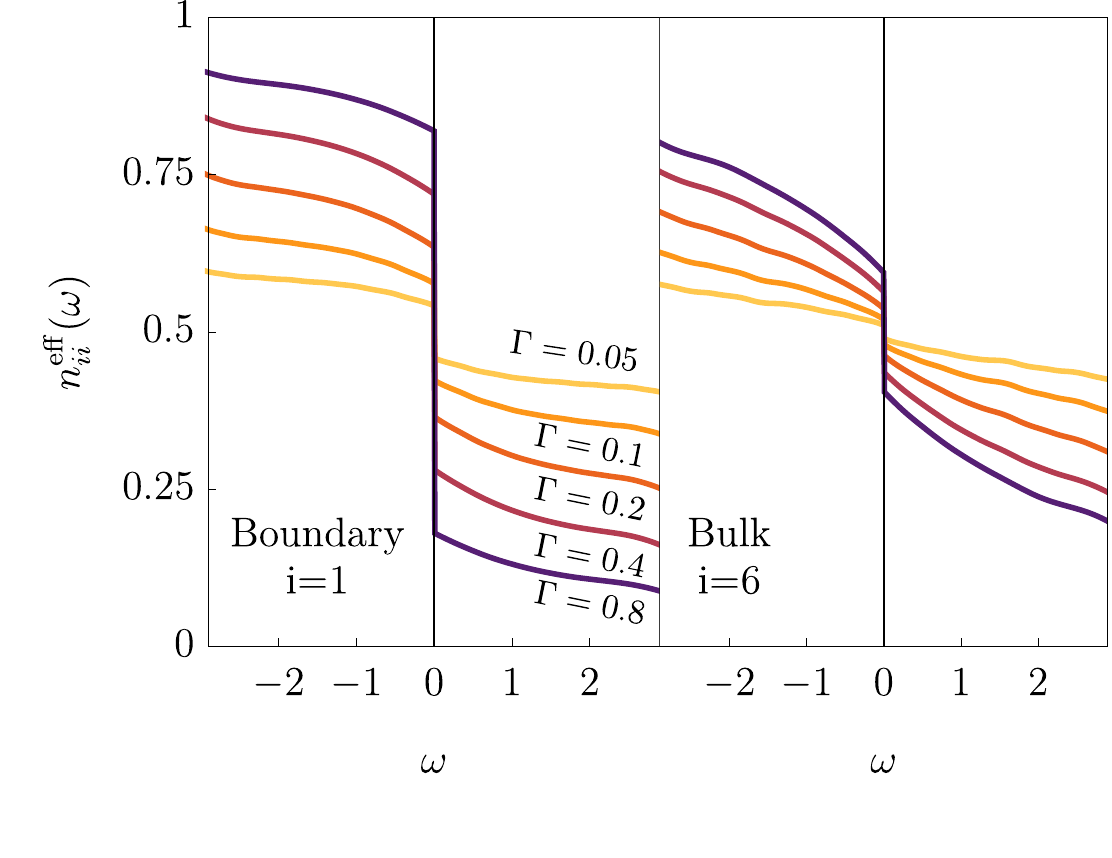}
    \caption{
        The effective distribution function at the boundary (left panel) and in the bulk (right panel) in equilibrium ($\mu_\mathrm{L,R}=0$) for $U=1$, $N=12$, and different reservoir couplings $\Gamma$. The data was computed using an infinite-temperature cutoff scheme. The exact result is given by Eq.~(\ref{eq:fluc_dis2}). The physical reservoirs `cool' the system towards zero temperature only at strong couplings and close to the boundary.  }
    \label{fig:noBiasVarGamComp}
\end{figure}

In equilibrium, the fluctuation-dissipation theorem holds [see Eq.~\eqref{eq:fluc_dis}]. The effective (non-equilibrium) distribution function defined in Eq.~\eqref{eq:fluc_dis_eff} should therefore reduce to the zero-temperature Fermi-Dirac distribution function,
\begin{equation}\label{eq:fluc_dis2}
n^\text{eff}_{ij}(\omega)=n(\omega)\delta_{i,j}=\theta(-\omega)\delta_{i,j}.
\end{equation}
We now test how well the FDT is preserved within our approximate FRG approach; the results are summarized in Fig.~\ref{fig:noBiasCutoffComp}. The left panel shows the \((1,1)\)-component of the effective distribution function. We find that a zero-temperature cutoff (\(T_\mathrm{cut}=\mu_\mathrm{cut}=0\)) preserves the FDT (up to numerical errors associated with integration routines). An infinite-temperature cutoff (\(T_\mathrm{cut}=\infty\)), however, introduces artificial heating: It leads to a significant (de)population of states (below) above the Fermi level and severely violates the FDT even at the end of the flow. In Fig.~\ref{fig:noBiasVarGamComp}, we illustrate how this artificial heating depends on the reservoir-coupling $\Gamma$ and on the distance to the boundary. A strong coupling to the physical zero-temperature reservoirs reduces this artificial heating substantially; the distribution function evolves towards a Fermi-Dirac distribution as $\Gamma$ is increased. This `cooling' effect, however, is local and almost absent in the bulk (see the right panel of Fig.~\ref{fig:noBiasVarGamComp}).

\begin{figure}[t]
    \includegraphics[width=\columnwidth]{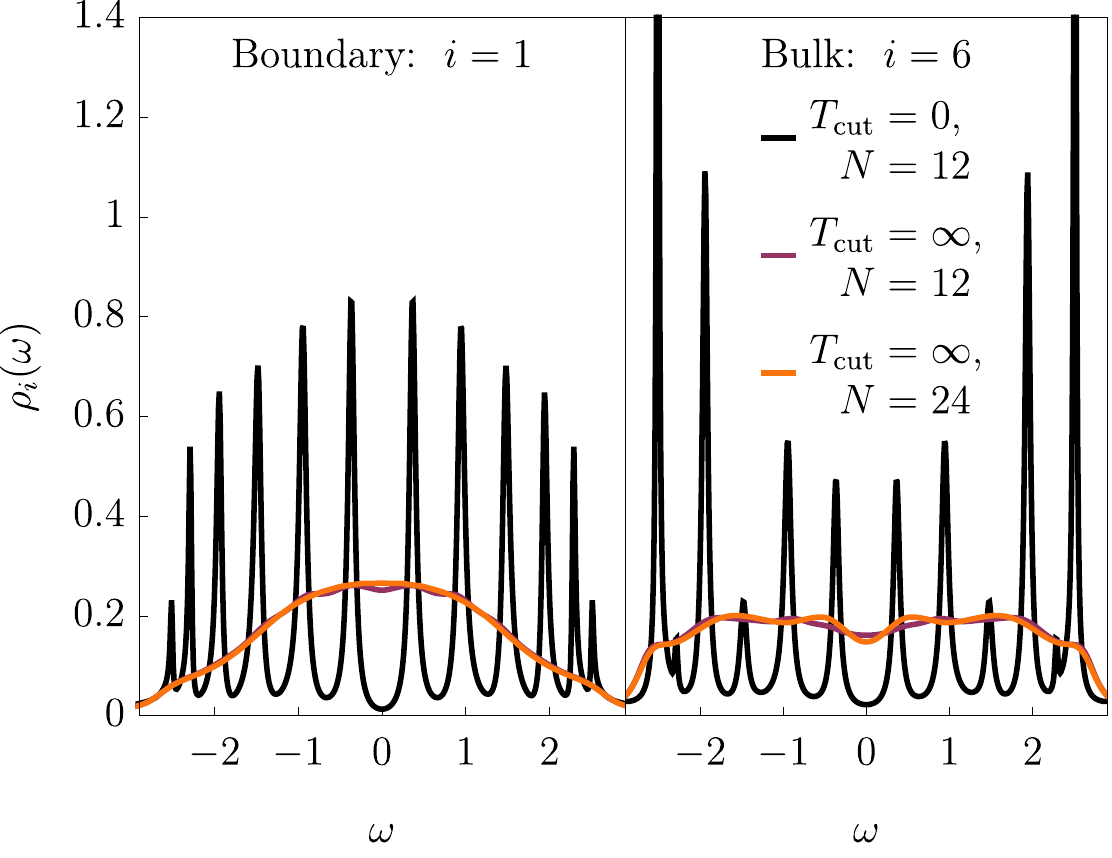}
    \caption{
        The local density of states at the boundary (left panel) and in the bulk (right panel) in equilibrium ($\mu_\mathrm{L,R,cut}=0$)  for \(\Gamma=0.2\) and \(U=1\). The various lines show data obtained for different cutoff temperatures $T_\tn{cut}$ and system sizes $N$. 
        In a Luttinger liquid, the local density of states is expected to vanish at the Fermi surface. In contrast, our infinite-temperature cutoff introduces inelastic scattering and yields a smooth LDOS (see also Fig.~\ref{fig:noBiasCutoffComp}).
    }
    \label{fig:noBiasCutoffCompDOS}
\end{figure}

\subsubsection{Scattering induced by hot reservoirs}

Next, we investigate the imaginary part of the self-energy \(\Im(\Sigma^\mathrm{adv}_{ii})(\omega)\). This quantity roughly measures the generation of inelastic scattering at site $i$. Results obtained for $T_\tn{cut}=0$ as well as $T_\tn{cut}=\infty$ are shown in the right panel of Fig.~\ref{fig:noBiasCutoffComp}.  In equilibrium, it is well-understood\cite{giamarchi2003quantum,Samokhin1998} that no additional inelastic scattering should be generated close to the Fermi-edge, \(\Im(\Sigma^\mathrm{adv}_{ii})(\omega=0)=0\). The zero-temperature cutoff reproduces this result. The infinite-temperature cutoff, however, artificially introduces such processes via the flow. This problem exacerbates for larger system as the influence of the physical coupling on the center of the chain decreases.

\subsubsection{Density of states}
Finally, we study the local density of states (see Fig.~\ref{fig:noBiasCutoffCompDOS}). While we cannot reproduce hallmarks of Luttinger liquid physics such as critical power laws\cite{giamarchi2003quantum} for small systems, we clearly find that an infinite-temperature cutoff yields a smooth LDOS even for \(N=12\). This again relates to the unphysical generation of an inelastic scattering length scale (even at the Fermi surface) that is comparable to (or smaller than) the system size. The zero-temperature cutoff does not yield a smooth LDOS. This is another demonstration of the cutoff-dependence in a physical quantity within our FRG approach.

\subsubsection{Summary}
We have demonstrated that our FRG results feature a strong cutoff-dependence in equilibrium. Since it is a priori unclear what cutoff scheme to employ away from this limit, our approach seems unsuitable in its present form. One loophole, however, remains: Thermal equilibrium is (counter-intuitively) the most challenging one situation, both from a numerical perspective (integrals become sharply-peaked) and potentially fundamentally (e.g, a delicate interplay of collective phenomena leads to the suppression of scattering around the Fermi-level). Therefore, we will now analyze whether or not the detrimental cutoff-dependence still shows up in non-equilibrium.

\subsection{Results at finite bias}
\label{ssec:bias_results}

Throughout this section, we drive the system out of equilibrium by applying a finite bias voltage \(\mu_\mathrm{L}=-\mu_\mathrm{R}=1\). We will compare data obtained using three different cutoff schemes: i) $T_\tn{cut}=\mu_\tn{cut}=0$, ii) $T_\tn{cut}=0,\mu_\tn{cut}=\mu_\tn{R}=-1$, and iii) $T_\tn{cut}=\infty$ (the choice of the chemical potential $\mu_\tn{cut}$ is irrelevant in the latter case). We again emphasize that the physical reservoirs are always held at zero temperature $T_\tn{L}=T_\tn{R}=0$.

\begin{figure}[t]
    \includegraphics[width=\columnwidth,clip]{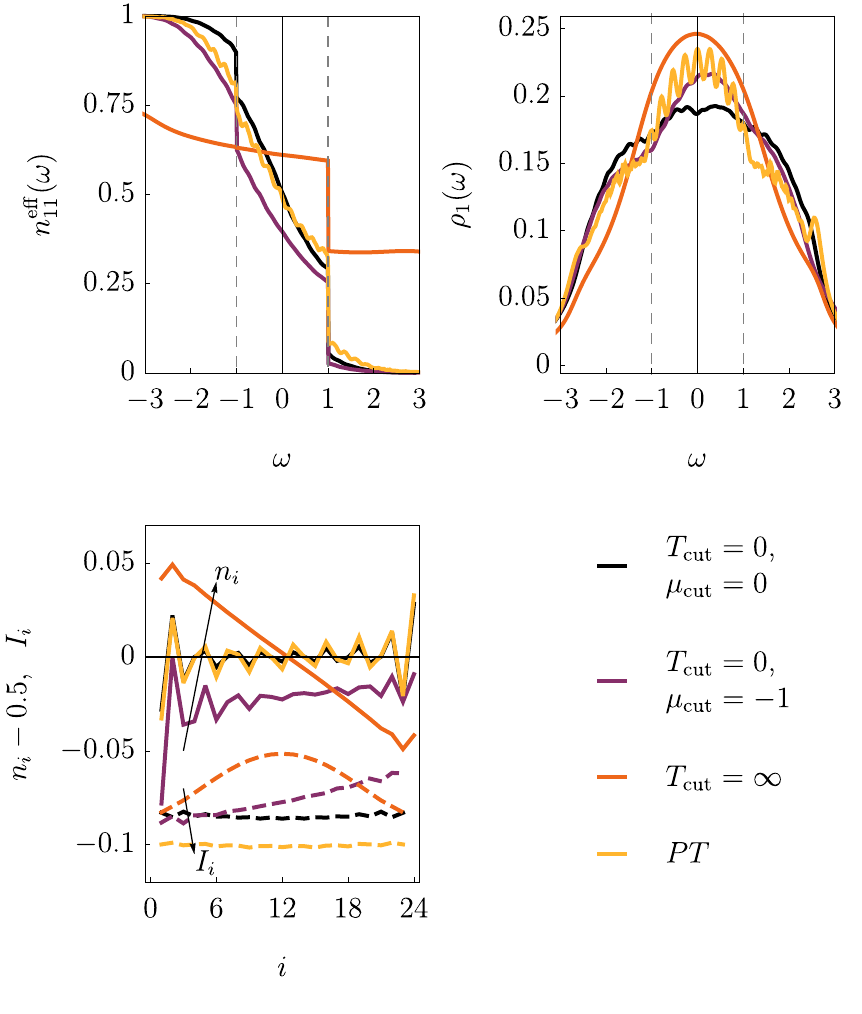}
    \caption{FRG results for a chain of $N=24$ sites with \(\Gamma=0.2\) and \(U=1\) obtained using different cutoff schemes. The system is driven out of equilibrium by a bias voltage $\mu_\text{L,R}=\pm1$. The second-order perturbation-theory result is shown for comparison (PT). \emph{Left panel}: The effective distribution function \eqref{eq:fluc_dis_eff} at the boundary. For small $\Gamma$ and $U$, one expects a piecewise-constant function with two steps of height \(1/2\) at the chemical potentials of the reservoirs. \cite{severindr}
        \emph{Center panel}: The local density of  states at the boundary. \emph{Right panel}: The occupation of the individual sites (solid) and the local current (dashed).
    }
    \label{fig:biasCutoffComp}
\end{figure}

\begin{figure}[t]
    \includegraphics[width=\columnwidth,clip]{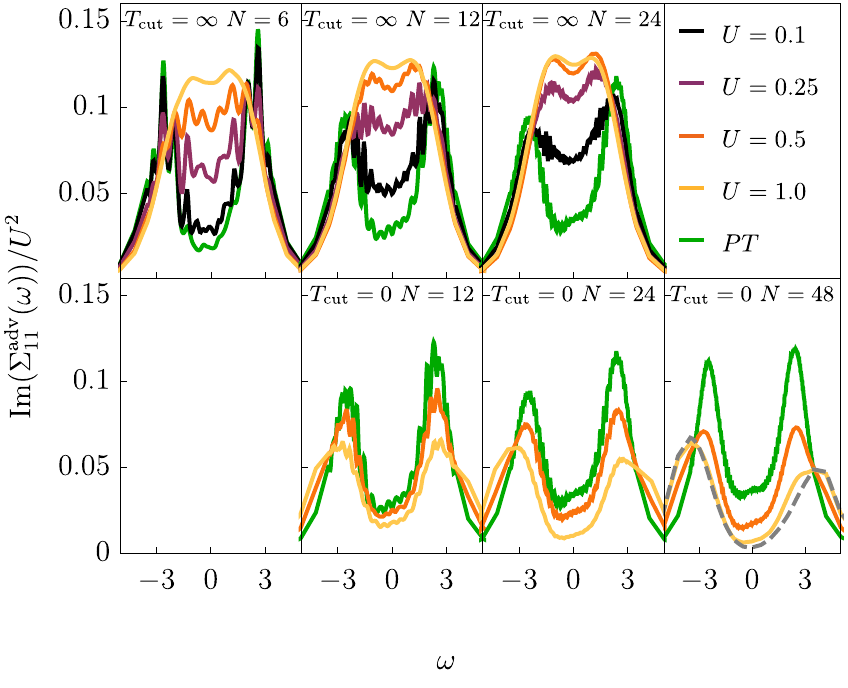}
    \caption{FRG results for the imaginary part of self-energy at the boundary of a chain with $\Gamma=0.2$, \(\mu_\mathrm{L,R}=\pm 1\), and for various values of $U$. The upper (lower) row shows data obtained using an infinite-temperature (zero-temperature) cutoff with $\mu_\mathrm{cut}=0$. The columns contain different system sizes $N=6,12,24,48$ (the dashed gray line in the bottom-right panel was calculated for $U=1$ with \(N=60\)). With increasing $N$, the dependence of the results on the cutoff becomes more pronounced. This illustrates that our FRG scheme is insufficient to reliably address the out-of-equilibrium properties of large, interacting systems.}
    \label{fig:biasCutoffCompSig}
\end{figure}

\subsubsection{Cutoff-dependence of physical observables}
In Fig.~\ref{fig:biasCutoffComp}, we show the effective distribution function, the local density of states, the local occupation number, and the local current for a system with \(N=24\) sites. FRG results were obtained using three different cutoff schemes; second-order perturbation-theory is included for comparison.

We find that all cutoffs yield an effective distribution function \(n_{11}^\mathrm{eff}(\omega)\) that breaks inversion symmetry, implying that the distribution function is not uniform throughout the chain (see the upper left panel of Fig.~\ref{fig:biasCutoffComp}).
This is in line with general expectations and remedies a shortcoming of a first order FRG approach.\cite{Jakobs2007} Unfortunately, the distribution function generally features a strong cutoff-dependence. 

The local density of states at the boundary is shown in the upper right panel of Fig.~\ref{fig:biasCutoffComp}.
To reduce finite-size effects, we introduce an artificial broadening via
\begin{equation}
    \rho_i(\omega)=-\frac{1}{\pi}\Im\left\{ \frac{1}{\left[ G^\mathrm{ret}(\omega)\right]^{-1} +\I \Gamma_\mathrm{smear}} \right\}_{ii},\ \Gamma_\mathrm{smear}=0.2.
\end{equation}
For a cutoff with \(T_{\rm cut}=\mu_{\rm cut}=0\), this quantity  starts to show a cusp at \(\omega=0\)  as well as shoulders at \(\omega=\mu_\mathrm{L}\) and \(\omega=\mu_\mathrm{R}\). Such features are expected in the ground state of a Luttinger liquid at \(\mu=0,\mu_\mathrm{L}$ and $\mu_\mathrm{R}\), respectively. They hint towards the survival of some of the ground-state Luttinger liquid physics even at finite bias. However, these features are not present in the data obtained using the other cutoff schemes; their appearance is uncontrolled.

The occupations (see the lower panel of Fig.~\ref{fig:biasCutoffComp})
\begin{equation}
    n_i=\left\langle c_i^\dagger c_i\right\rangle=\frac{1}{2}-\frac{\I}{2}\int\frac{\dOp\omega}{2\pi} G^\mathrm{K}_{ii}(\omega)
\end{equation}
show a gradient-type behavior within the infinite-temperature cutoff, while in the other schemes Friedel oscillations dominate. The stationary-state current 
\begin{equation}
    I_i=\Re \int \frac{\dOp \omega}{2\pi} G^\mathrm{K}_{i(i+1)}(\omega)
\end{equation}
is conserved within perturbation theory, while the FRG schemes violate particle-number conservation to $\mathcal{O}(U^3)$. This is especially severe for \(T_\mathrm{cut}=\infty\), which leads to a strong suppression of the current in the middle of the chain.

In a nutshell, the FRG results for physical quantities feature a strong cutoff-dependence out-of-equilibrium.

\subsubsection{Inelastic scattering and scaling}
As in equilibrium, it is insightful to analyze the imaginary part of the self-energy. Results are shown in Fig.~\ref{fig:biasCutoffCompSig} for two different cutoff schemes; perturbation theory is included for comparison. We find that a cutoff with \(T_\mathrm{cut}=\mu_\mathrm{cut}=0\) (with $T_\mathrm{cut}=\infty$)   consistently underestimates (overestimates) the amount of inelastic processes compared to perturbation theory.
As such, this is not necessarily problematic. However, the difference between perturbation theory and
the FRG data increases with increasing $N$. In that sense, the system does not behave perturbatively, as `secular' terms in the system size $N$ such as $U^3N$ arise. Note that this strong cutoff dependence does not only occur in the bulk (i.e., far from the cold physical reservoirs) but also right at the boundary of the interacting system. Thus, our second-order FRG scheme is not suited to study systems out of thermal equilibrium at least with the (reservoir) cutoffs employed.


\begin{table*}[t]
    \begin{tabular}{l||p{1cm}|p{4.5cm}|p{7cm}}
        & ret & adv & K\\
        \hline
        \hline
        ret & \(0\) & \(\pm Q_{q_1}\otimes Q^\dagger_{q_2}f_0(\pm\lambda_{q_1}, \lambda^*_{q_2}-\omega)\) & \(\pm Q_{q_1} \otimes Q_{q_2} \eta_\alpha f_1(\pm\lambda_{q_1},\lambda_{q_2}-\omega, \mu_\alpha)\)\newline\(\mp Q_{q_1} \otimes\eta_\alpha Q^\dag_{q_2}  f_1(\pm\lambda_{q_1},\lambda^*_{q_2}-\omega, \mu_\alpha)\) \\
        \hline
        adv & & \(0\) &\(\pm Q^\dag_{q_1} \otimes Q_{q_2} \eta_\alpha f_1(\pm\lambda^*_{q_1},\lambda_{q_2}-\omega, \mu_\alpha)\)\newline\(\mp Q^\dag_{q_1} \otimes\eta_\alpha Q^\dag_{q_2}  f_1(\pm\lambda^*_{q_1},\lambda^*_{q_2}-\omega, \mu_\alpha)\) \\
        \hline
        K& & &
        \(\pm Q_{q_1}\eta_{\alpha_1} \otimes Q_{q_2} \eta_{\alpha_2} f_2(\pm\lambda_{q_1},\lambda_{q_2}-\omega, \mu_{\alpha_1}, \mu_{\alpha_2})\)
        \newline
        \(\mp Q_{q_1}\eta_{\alpha_1} \otimes  \eta_{\alpha_2}Q^\dagger_{q_2} f_2(\pm\lambda_{q_1},\lambda^*_{q_2}-\omega, \mu_{\alpha_1}, \mu_{\alpha_2})\)
        \newline
        \(\mp \eta_{\alpha_1}Q^\dagger_{q_1} \otimes Q_{q_2} \eta_{\alpha_2} f_2(\pm\lambda^*_{q_1},\lambda_{q_2}-\omega, \mu_{\alpha_1}, \mu_{\alpha_2})\)
        \newline
        \(\pm \eta_{\alpha_1}Q^\dagger_{q_1} \otimes  \eta_{\alpha_2}Q^\dagger_{q_2} f_2(\pm\lambda^*_{q_1},\lambda^*_{q_2}-\omega, \mu_{\alpha_1}, \mu_{\alpha_2})\)
    \end{tabular}
\caption{Analytical expressions for \(\int \dOp \Omega g^\mathrm{row}(\pm\Omega) g^\mathrm{col}(\Omega+\omega)\). For readability, we omit all summations as well as the single-particle indices. Both are to be understood in analogy to Eq.~\eqref{eq:exampleTwoGF}. The missing entries of the table can be obtained by using \(\int \dOp \Omega g^\mathrm{row}(\pm\Omega) g^\mathrm{col}(\Omega+\omega)=\int \dOp \Omega g^\mathrm{col}(\pm\Omega)g^\mathrm{row}(\Omega\mp \omega) \).}
\label{tab:twgf}
\end{table*}

\section{Conclusion}

We developed a second-order implementation  of the Keldysh functional renormalization group to study out-of-equilibrium quantum wires attached to non-interacting reservoirs. Our key idea is to simplify the flow equation of the two-particle vertex by neglecting its own feedback as well as the feedback of the self-energy. This approach is correct to second order in the interaction but still contains an infinite resummation of higher-order terms (since the flow of the self-energy is solved in full). By combining semi-analytic solution techniques with massive MPI parallelization, we treated system of up to 60 lattice sites.

Within the FRG, we employed a so-called reservoir cutoff, which is physical, easy to implement, numerically-efficient, and which has proven to provide good results in other setups.\cite{Karrasch2010,Jakobs2010a,Jakobs2010b,Kennes2012} Since one can vary the temperature and the chemical potential of the auxiliary reservoirs, our approach in fact encompasses a whole class of cutoffs. This has the distinct advantage that one can explicitly analyze whether or not the results are indeed independent of the particular choice of the RG procedure.

As a prototypical model, we studied a one-dimensional tight-binding chain with nearest-neighbor interactions that is weakly coupled to left and right reservoirs. We computed effective distribution functions, the local density of states, and the steady-state current and demonstrated that all of these quantities depend strongly on the choice of the cutoff. Exact results (such as the fluctuation-dissipation theorem) are available in equilibrium and can serve as a benchmark for the FRG data. In non-equilibrium, there is no physically-motivated cutoff; moreover, secular higher-order terms appear which are only partly included in a our approach. This demonstrates that our second-order FRG scheme is highly-demanding but still inadequate to study interacting quantum wires out of equilibrium.

A different cutoff scheme or, if possible, a more thorough treatment of the higher-order vertices might yield better results and are intriguing avenues of future research. Furthermore, for systems where physical inelastic processes limit the coherence length, our method is still expected to provide insights on how small interactions modify their behavior through heating.

\section*{Acknowledgments} DMK was funded by the Deutsche Forschungsgemeinschaft (DFG, German Research Foundation) under Germany's Excellence Strategy - Cluster of Excellence Matter and Light for Quantum Computing (ML4Q) EXC 2004/1 - 390534769. We acknowledge support from the Max Planck-New York City Center for Non-Equilibrium Quantum Phenomena. CKa and CKl acknowledge support by the Deutsche Forschungsgemeinschaft through the Emmy Noether program (KA 3360/2-2). CKa acknowledges support by the `Niedersächsisches Vorab' through `Quantum- and Nano-Metrology (QUANOMET)' initiative within the project P-1. 

\appendix

\section{Perturbative two-particle vertex}\label{ch:twoGF}
In the main text, we discussed how to treat the flow of the two-particle vertex on a perturbative level.
In order to obtain an efficient algorithm, we relied on analytic expressions for all two-Green's-function integrals (see Sec.~\ref{sec:fin_ala_vert}). In this appendix, we are going to discuss in more detail how these can be obtained.

Using Eq.~\eqref{eq:grDecomp} for the retarded and advanced component of the Green's function as well as Eqs.~\eqref{eq:sylvester2} and \eqref{eq:sylvesterIndiRes} for the Keldysh component, the fermionic degrees of freedom can be decoupled from the frequency dependence. As an example, Eq.~\eqref{eq:exampleTwoGF} shows how to decompose the retarded-Keldysh two-Green's function integral. The remaining frequency integrals are all of one of three types:
\begin{equation}
    \begin{split}
        f_0(a,b)&=\int d\Omega \frac{1}{\Omega-a} \frac{1}{\Omega-b},\\
        f_1(a,b,\mu)&=\int d\Omega \frac{1}{\Omega-a} \frac{1}{\Omega-b} \sgn(\Omega-\mu),\\
        f_2(a,b,\mu,\mu')&=\int d\Omega \frac{1}{\Omega-a} \frac{1}{\Omega-b} \sgn(\Omega-\mu)\sgn(\Omega-\mu').\\
    \end{split}
\end{equation}

In order to set up an efficient notation, we introduce the indefinite integral
\begin{equation}
    \begin{split}
        h(\Omega, a,b)&=\int d\Omega \frac{1}{\Omega-a} \frac{1}{\Omega-b}\\
    &=\begin{cases}
        \frac{1}{a-b}\left(\ln(\Omega-a)-\ln(\Omega-b)\right)& a\neq b\\
        \frac{1}{a-\Omega}& a= b,
    \end{cases}
    \end{split}
\end{equation}
with the limiting behavior
\begin{equation}
    \begin{split}
    &h(\infty,a,b)=0,\\
    &h(-\infty,a,b)=\begin{cases}
        -\left[ \sgn(\Im a)- \sgn(\Im b)\right]\frac{\I\pi}{a-b}&a\neq b\\
        0                                                      & a= b.\\
    \end{cases}
    \end{split}
\end{equation}
We placed the branch cut along the negative real axis. The three integrals $f_{0,1,2}$ are then given by
\begin{equation}
    \begin{split}
        f_0(a,b)&=h(\infty,a,b)-h(-\infty,a,b)\\
        &=
            \begin{cases}
                0\hspace{1cm} &\sgn(\Im a)=\sgn(\Im b)\\
                2\pi \I \frac{1}{a-b} &\Im a>0,\ \Im b<0\\
                -2\pi \I \frac{1}{a-b} &\Im a<0,\ \Im b>0,
            \end{cases}
    \end{split}
\end{equation}
\begin{equation}
    \begin{split}
        f_1(a,b,\mu)&=h(\infty,a,b)-h(\mu,a,b)\\
                    &-\left[h(\mu,a,b)-h(-\infty,a,b)\right]\\
                    &=-2h(\mu,a,b)-\left[ \sgn(\Im a)- \sgn(\Im b)\right]\frac{\I\pi}{a-b},\\[2ex]
    \end{split}
\end{equation}
and
\begin{equation}
    \begin{split}
        f_2(a,b,\mu,\mu')
        &=\begin{cases}
                f_0(a,b) & \mu=\mu'\\
                -2h(\mu',a,b)+2h(\mu,a,b)& \mu<\mu'\\
                +\left[ \sgn(\Im a)- \sgn(\Im b)\right]\frac{\I\pi}{a-b} &
            \end{cases}
    \end{split}
\end{equation}
where for convenience we set \(0/0=0\).

With these, a full table of the two-Green's function integrals can be provided (as shown in table~\ref{tab:twgf}).\\

\bibliography{ref}

\end{document}